\documentclass[fleqn,usenatbib]{mnras}

\usepackage{newtxtext,newtxmath}

\usepackage[T1]{fontenc}
\usepackage{ae,aecompl}

\usepackage{listings}
\usepackage{xcolor}

\usepackage{graphicx}	
\usepackage{amsmath}	
\usepackage{hyperref}
\usepackage{natbib}
\usepackage[caption=false]{subfig}
\usepackage{ragged2e}
\usepackage{placeins}
\usepackage{bigints} 
\usepackage{physics}
\usepackage{xcolor}
\usepackage{cleveref}
\usepackage[export]{adjustbox}
\usepackage{longtable}
\usepackage{xtab}

\usepackage{array}
\usepackage{multirow}
\usepackage{hhline,booktabs}
\usepackage{makecell}

\usepackage{scalerel,tikz}
\usetikzlibrary{svg.path}
\definecolor{orcidlogocol}{HTML}{A6CE39}
\tikzset{orcidlogo/.pic={
 \fill[orcidlogocol] svg{M256,128c0,70.7-57.3,128-128,128C57.3,256,0,198.7,0,128C0,57.3,57.3,0,128,0C198.7,0,256,57.3,256,128z};
 \fill[white] svg{M86.3,186.2H70.9V79.1h15.4v48.4V186.2z}
 svg{M108.9,79.1h41.6c39.6,0,57,28.3,57,53.6c0,27.5-21.5,53.6-56.8,53.6h-41.8V79.1z M124.3,172.4h24.5c34.9,0,42.9-26.5,42.9-39.7c0-21.5-13.7-39.7-43.7-39.7h-23.7V172.4z}
 svg{M88.7,56.8c0,5.5-4.5,10.1-10.1,10.1c-5.6,0-10.1-4.6-10.1-10.1c0-5.6,4.5-10.1,10.1-10.1C84.2,46.7,88.7,51.3,88.7,56.8z};
}}
\newcommand\orcidicon[1]{\href{https://orcid.org/#1}{\mbox{\scalerel*{
\begin{tikzpicture}[yscale=-1,transform shape]
\pic{orcidlogo};
\end{tikzpicture}
}{|}}}}

\def\sun{\odot}

\defcitealias{Mohapatra2021b}{M21}
\defcitealias{Li2020ApJ}{Li20}
\defcitealias{Wang2021MNRAS}{W21}

\title[ICM gas kinematics]{Velocity structure functions in multiphase turbulence: interpreting kinematics of H$\alpha$ filaments in cool core clusters}

\author[Mohapatra et al]{
Rajsekhar Mohapatra$^{\orcidicon{0000-0002-1600-7552}\,1}$\thanks{E-mail: rajsekhar.mohapatra@anu.edu.au (RM)},
Mrinal Jetti$^{\orcidicon{0000-0003-0287-3246}\,2}$\thanks{E-mail:mrinaljetti@gmail.com  (MJ)},
Prateek Sharma$^{\orcidicon{0000-0003-2635-4643}\,3}$\thanks{E-mail: prateek@iisc.ac.in (PS)} and
Christoph Federrath$^{\orcidicon{0000-0002-0706-2306}\,1,4}$\thanks{E-mail: christoph.federrath@anu.edu.au (CF)} 
\\
$^{1}$Research School of Astronomy and Astrophysics, Australian National University, Canberra, ACT~2611, Australia\\
$^{2}$ Department of Aerospace Engineering, Indian Institute Of Technology, Chennai, Tamil Nadu 600036, India\\
$^{3}$ Department of Physics, Indian Institute of Science, Bangalore, KA 560012, India\\
$^{4}$Australian Research Council Centre of Excellence in All Sky Astrophysics (ASTRO3D), Canberra, ACT~2611, Australia\\
}

\date{Accepted XXX. Received YYY; in original form ZZZ}

\pubyear{2021}

\begin{document}
\label{firstpage}
\pagerange{\pageref{firstpage}--\pageref{lastpage}}
\maketitle

\begin{abstract}
The central regions of cool-core galaxy clusters harbour multiphase gas, with gas temperatures ranging from $10$~$\mathrm{K}$--$10^7$~$\mathrm{K}$. Feedback from active galactic nuclei (AGNs) jets prevents the gas from undergoing a catastrophic cooling flow. However, the exact mechanism of this feedback energy input is unknown, mainly due to the lack of velocity measurements of the hot phase gas.
However, recent observations have measured the velocity structure functions ($\mathrm{VSF}$s) of the cooler molecular ($\sim10$~$\mathrm{K}$) and H$\alpha$ filaments ($\sim10^4$~$\mathrm{K}$) and used them to indirectly estimate the motions of the hot phase. In the first part of this study, we conduct high-resolution ($384^3$--$1536^3$ resolution elements) simulations of homogeneous isotropic subsonic turbulence, without radiative cooling. We analyse the second-order velocity structure functions ($\mathrm{VSF}_2$) in these simulations and study the effects of varying spatial resolution, the introduction of magnetic fields, and the effect of projection along the line of sight (LOS) on it. In the second part of the study, we analyse high-resolution ($768^3$ resolution elements) idealised simulations of multiphase turbulence in the intracluster medium (ICM) from \cite{Mohapatra2021b}. 
We compare the $\mathrm{VSF}_2$ for both the hot ($T\sim10^7$~$\mathrm{K}$) and cold ($T\sim10^4$~$\mathrm{K}$) phases and find that their amplitude depends on the density contrast between the phases. They have similar scaling with separation, but introducing magnetic fields steepens the $\mathrm{VSF}_2$ of only the cold phase. We also find that projection along the LOS steepens the $\mathrm{VSF}_2$ for the hot phase and mostly flattens it for the cold phase.
\end{abstract}




\section{Introduction}\label{sec:introduction}
The intracluster medium (ICM) refers to the hot X-ray emitting plasma in galaxy clusters. In the central regions of relaxed clusters, the ICM is denser and often has short cooling times ($\sim$ a few $100$~$\mathrm{Myr}$). These regions are often multiphase and consist of hot ionised regions at $10^7$~$\mathrm{K}$ that are co-spatial with the cooler atomic phase at $\sim10^4$~$\mathrm{K}$ \citep{hu1992,Conselice2001AJ,mcdonald2012,Lakhchaura2018MNRAS} and the molecular phase at $\sim 10$~$\mathrm{K}$ \citep{edge2001,Pulido2018ApJ,Rose2019MNRAS,Vantyghem2019ApJ,Vantyghem2021ApJ}. The cool and dense core is supposed to undergo a runaway cooling flow, till it forms molecular gas \citep{fabian1994cooling}. However, feedback through active galactic nuclei (AGNs) jets from the central supermassive blackholes (SMBH) of the clusters injects energy back into the ICM. This energy heats up the core and prevents a catastrophic cooling flow. 

Many different mechanisms have been proposed to be a channel of energy transfer from AGN jets to the ICM. These include sound waves \citep{Tang2018MNRAS,Bambic2019ApJ}, shocks \citep{Yang2016ApJ},  heating by turbulent dissipation \citep{zhuravleva2014turbulent}, heating by turbulent mixing \citep{banerjee2014turbulence,Hillel2016MNRAS,Fujita2020MNRAS}, internal gravity waves \citep{Zhang2018}, cosmic rays (CRs) \citep{Pfrommer2013ApJ,Ruszkowski2018ApJ}, etc. Measuring velocities of the ICM plasma is important to  constrain these different heating methods and for the physical understanding of the ICM. 

Many observational studies have used both direct and indirect methods to estimate turbulent velocities of the ICM, summarised in \cite{Simionescu2019SSRv}. Direct measurements of hot phase gas velocities are difficult due to large thermal velocities of the plasma. The Hitomi space telescope \citep{hitomi2016} directly measured velocities of the hot-phase gas by resolving the line-broadening of Fe\texttt{xxv} and Fe\texttt{xxvi} lines and revealed low levels of turbulence. Recently, observations have obtained indirect estimates of the level of turbulent velocities from X-ray brightness fluctuations \citep{zhuravleva2014turbulent,zhuravleva2018} and thermal Sunyaev-Zeldovich (tSZ) fluctuations \citep{Zeldovich1969,Sunyaev1970Ap&SS,khatri2016}, which represent line of sight (LOS) electron pressure fluctuations. \citet{Li2020ApJ} (hereafter \citetalias{Li2020ApJ})  have used optical data to obtain turbulent velocities of ICM gas in the atomic filaments in several nearby clusters. They have analysed the first-order structure functions of LOS velocity ($\mathrm{VSF}_1^{\mathrm{LOS,obs}}$) and find them to be steeper than expected from Kolmogorov turbulence theory \citep[][K41]{kolmogorov1941dissipation}. Velocity structure functions ($\mathrm{VSF}$s) and spatial power spectra are useful diagnostics since they represent the variation of velocity with scale. Using $\mathrm{VSF}_1^{\mathrm{LOS,obs}}$, \citetalias{Li2020ApJ} show that the driving scale of turbulence in their sample of clusters corresponds to the size of AGN-driven bubbles. 

Recent observational studies have also looked at structure functions and power spectra corresponding to the interstellar medium (ISM) and the intergalactic medium (IGM). \cite{Ha2021ApJ} have studied the $\mathrm{VSF}_1$ of different star groups in the Orion complex. \cite{Marchal2021arXiv} have studied the density power spectra of the multiphase high velocity cloud complex C. \cite{Qian2018ApJ,Xu2020MNRAS} have studied the $\mathrm{VSF}_2$ of the Taurus molecular cloud where they report a transition from K41 scaling to Burgers scaling \cite{Burgers1948171} of the $\mathrm{VSF}_2$. \cite{Xu2020ApJ} have looked at the structure function of dispersion measures of fast radio bursts to study the density variations due to IGM turbulence.

Motivated by these observational techniques, recent numerical studies have also studied the $\mathrm{VSF}$s or velocity power spectra in similar multiphase ICM environments. In \cite{Mohapatra2019}, we looked at the power spectra of density and velocity and studied the effects of radiative cooling and thermodynamics of the ICM in idealised local box simulations of size $(40$~$\mathrm{kpc})^3$. \cite{Hillel2020ApJ} have analysed the jet-ICM interaction in a $(40$~$\mathrm{kpc})^3$ octant of a galaxy cluster core. They argue that bubbles inflated by AGN jets can drive turbulence over a large range of scales, which sets the slope of the first-order velocity structure function $(\mathrm{VSF}_1)$.
\cite{Wang2021MNRAS}  (hereafter \citetalias{Wang2021MNRAS}) have simulated a galaxy cluster in a $(1$~$\mathrm{Mpc})^3$ cube. They have studied the $\mathrm{VSF}_1$ of the cold and hot phases and the coupling between them. They show that the gas in the cold phase is magnetically supported (see also \citealt{Nelson2020MNRAS}) and follows a steeper $\mathrm{VSF}_1$ compared to the hot phase, which is expected from free-falling clouds. 

However, \cite{Hillel2020ApJ} use an adaptive mesh refinement (AMR) grid for their simulations. This sets a variable inertial range of turbulence (due to varying numerical viscosity), which would make it difficult to analyse the exact slope of the structure functions. Their simulations are evolved for a short time and hence  do not have multiphase gas. The $\mathrm{VSF}_1$ of the hot and cold phases could be different, as seen in \citetalias{Wang2021MNRAS}. Cooling and the presence of cold phase gas can also affect the scaling of velocity power spectra at small scales, as seen in \cite{Mohapatra2019}. On the other hand, \citetalias{Wang2021MNRAS} use static mesh refinement, with higher resolution towards the denser central region of the cluster. However, these also lead to the spatial variation of numerical viscosity, which can affect the shape of the $\mathrm{VSF}_1$. All three of these studies have also not looked into the effect of projection along the LOS on the  $\mathrm{VSF}_1$. \cite{Zuhone2016APJ} show that $\mathrm{VSF}_1^{\mathrm{LOS}}$ could be steeper than than the three-dimensional $\mathrm{VSF}_1$. Hence to understand the observational results and their implications, we need high resolution simulations of multiphase turbulence (with converged slopes of power spectra/$\mathrm{VSF}$s) in the ICM. 

This study is divided into two parts. In the first part, we conduct idealised simulations of homogeneous isotropic turbulence at subsonic rms Mach numbers ($\mathcal{M}$), without radiative cooling of the ICM. We study the slope of structure functions and their convergence with resolution. We also quantify the effects of magnetic fields and projection along the LOS on the $\mathrm{VSF}$s. In the second part, we analyse the high resolution multiphase ICM turbulence simulations from \cite{Mohapatra2021b} (hereafter \citetalias{Mohapatra2021b}). In these runs we study the $\mathrm{VSF}$s of the hot and cold phases separately. We compare 
$\mathrm{VSF}$s with and without magnetic fields and two different types of idealised thermal heating models (namely heating $\propto\rho^1$ or $\propto\rho^0$, where $\rho$ is the density of the gas).  We also analyse the effect of projection using different weightings along the LOS.

This paper is organised as follows. In \cref{sec:Methods}, we describe our setup for part 1 (without cooling runs). In \cref{sec:results-nocool}, we present the results from our idealised turbulence runs without cooling. Following this, in \cref{sec:results-cool}, we present the results from our analysis of multiphase simulations in \citetalias{Mohapatra2021b}. We discuss our results and compare them to results from other studies in \cref{sec:discussions}. In \cref{sec:caveats-future}, we discuss the caveats of our study and future prospects. We finally summarise and conclude in \cref{sec:Conclusion}.

\section{Methods}\label{sec:Methods}

\subsection{Model equations}\label{subsec:model_equations}
We model the ICM as a fluid using the compressible magnetohydrodynamic (MHD) equations. The gas follows an ideal gas equation of state. Our model equations are listed in section 2.1 of \citetalias{Mohapatra2021b} (numbered eq.~1a to 1f). For the first part of this study, we switch off radiative cooling and thermal heating. Thus, the energy equation becomes (instead of eq.~1c in \citetalias{Mohapatra2021b}):
\begin{equation}
    \frac{\partial E}{\partial t}+\nabla\cdot ((E+P^*)\mathbf{v}-(\mathbf{B}\cdot\mathbf{v})\mathbf{B})=\rho\mathbf{F}\cdot\mathbf{v},
	\label{eq:energy_nocool}
\end{equation}
where $\rho$ is the gas mass density, $\mathbf{v}$ is the velocity, $\mathbf{B}$ is the magnetic field, $P^*=P+\frac{\mathbf{B}\cdot\mathbf{B}}{2}$ is the sum of thermal and magnetic pressure, $P=\rho k_B T/(\mu m_p)$, $\mathbf{F}$ is the turbulent force per unit mass, $E$ 
is the total energy density, $\mu$ is the mean molecular mass, $m_p$ is the proton mass, $k_B$ is the Boltzmann constant, and $T$ is the temperature.

\subsection{Numerical setup}\label{subsec:numerical_setup}
We use a modified version of the FLASH code \citep{Fryxell2000,Dubey2008}, with the HLL5R Riemann solver \citep{Bouchut2007,Bouchut2010,Waagan2011} for solving the MHD equations. 
We use a uniformly spaced 3D Cartesian grid with $L_x=L_y=L_z=L=40$~$\mathrm{kpc}$, where $L$ is the box size. We use periodic boundary conditions along all three directions. Our default resolution is $768^3$ grid cells. In part 1, we also perform simulations with $384^3$ and $1536^3$ resolution elements for checking convergence. 

\subsection{Turbulence driving}\label{subsec:turb_driving}
We drive turbulence by using the stochastic Ornstein-Uhlenbeck (OU) process with a finite auto-correlation timescale \citep{eswaran1988examination,schmidt2006numerical,federrath2010}. We excite only large-scale modes in Fourier space ($1\leq\abs{\mathbf{k}}L/2\pi\leq3$, where $\mathbf{k}$ is the wave vector) and remove the divergent component. We inject power as a parabolic function of $\abs{\mathbf{k}}$, which peaks at $\abs{\mathbf{k}}=4\pi/L$, corresponding to half of the box-size. Turbulence at scales smaller than the driving scales ($\ell<L/3$) develops self-consistently through a nonlinear cascade. We refer the reader to section~2.1 of \citep{federrath2010} for more details of the turbulence driving method. 

\subsection{Initial conditions}\label{subsec:init_conditions}
We conduct new simulations only for part-1 of this study (turbulence without cooling), which provide a benchmark to compare against the more complex multiphase turbulence simulations. For consistency, we initialise all our simulations with the same conditions as in \citetalias{Mohapatra2021b}, with $T=4\times10^6$~$\mathrm{K}$, $n_e=0.086\,\mathrm{cm}^{-3}$. 
For our MHD runs, we set the initial magnetic field with equal mean and rms components. We set the initial plasma beta $\beta$ (ratio of thermal to magnetic pressure) to $100$, in line with observations \citep{Carilli2002ARA&A,Govoni2004IJMPD}. We set the mean field in the $z$ direction and the rms component as a power-law according to small-scale dynamo growth following the Kazantsev spectrum \citep{Kazantsev1968JETP}. The Fourier amplitude of the rms component $\mathrm{B}_k\propto k^{1.25}$, for $2\leq k \leq 20$.

\subsection{List of simulation models}\label{subsec:list_of_models}
We have conducted four new simulations without cooling for this study and analyse three high-resolution simulations with cooling from \citetalias{Mohapatra2021b}. We list all of them, including some of the relevant simulation parameters in \cref{tab:sim_params}. 

The without-cooling runs are listed in the upper part of the table and the with-cooling runs in the lower part, indicated by `f' in the label. Most of these runs use hydrodynamic equations (HD), by setting $\mathbf{B}=0$ throughout the domain. The runs with MHD are denoted by `mag' in the label. The default resolution is $768^3$, indicated by `HR' (high resolution) in the label. The $384^3$ and $1536^3$ resolution runs are labelled `LR' (low resolution) and `VHR' (very high resolution), respectively. We use these simulations for checking the convergence of our $\mathrm{VSF}$s.
For the with-cooling runs, the number after `f' denotes the parameter $f_{\mathrm{turb}}$, which is the ratio between the turbulent energy injection rate and the radiative cooling rate. All our runs have $f_{\mathrm{turb}}=0.10$. In these runs, `f0.10HR' and `f0.10magHR' have mass-weighted thermal heating with $Q_{\mathrm{mw}}\propto\rho$ and the `f0.10vwHR' run has $Q_{\mathrm{vw}}\propto\rho^0$, where $Q$ is the thermal heating rate density. We have described these methods in detail in section 2.2.2 of \citetalias{Mohapatra2021b}. 
We show the rms Mach number of the hot phase gas ($\mathcal{M}_{\mathrm{hot}}$) in column 5. For the without-cooling runs, this column just denotes the rms Mach number ($\mathcal{M}$). The hot phase is heated more in the volume weighted run, so even though the rms velocity in the hot phase is similar to the mass-weighted run, the Mach number is smaller.

\begin{table*}
	\centering
	\caption{Simulation parameters for different runs}
	\label{tab:sim_params}
	\resizebox{\textwidth}{!}{
		\begin{tabular}{lccccccc} 
			\hline
			Label & Resolution & $\mathrm{f}_{\mathrm{turb}}$ & Thermal heating &$\mathcal{M}_{\mathrm{hot}}$ & Magnetic fields  & Radiative cooling & $t_{\mathrm{end}}$ $(\mathrm{Gyr})$\\
			(1) & (2) & (3) & (4) & (5) & (6) & (7) & (8)\\
			\hline
			\textbf{Without cooling runs} \\
			LR & $384^3$ & NA & N & $0.26\pm0.01$ & N & N &  $1.042$\\
			HR & $768^3$ & NA & N & $0.27\pm0.01$ & N & N &  $1.042$\\
			magHR & $768^3$ & NA & N & $0.17\pm0.01$ & Y & N & $2.083$\\
			VHR & $1536^3$ & NA & N & $0.27\pm0.01$ & N & N &  $1.042$\\
			\hline
			\textbf{With cooling runs} (from \citetalias{Mohapatra2021b}) \\
			f0.10HR (fiducial) & $768^3$ & $0.10$ & mass-weighted & $0.91\pm0.05$ & N & Y & $1.003$\\
			f0.10vwHR & $768^3$ & $0.10$ & volume-weighted & $0.23\pm0.01$ & N & Y & $1.003$\\
			f0.10magHR & $768^3$ & $0.10$ & mass-weighted & $0.59\pm0.05$ & Y & Y & $1.003$\\
		\hline
			
	\end{tabular}}
	\justifying \\ \begin{footnotesize} Notes: Column~1 shows the simulation label. We list the without-cooling runs in upper half of the table and the with-cooling runs in the lower half. The default resolution of all the runs is $768^3$ cells, indicated by the label $\mathrm{HR}$. We use $\mathrm{LR}$ and $\mathrm{VHR}$ to denote $384^3$ and $1536^3$ cells, respectively (also shown in column~2). For our second set of runs (turbulence with cooling), the number following `f' in the label denotes $\mathrm{f}_{\mathrm{turb}}$ (also shown in column~3), which is the ratio of turbulent energy input rate to the radiative cooling rate. Column~4 lists the type of thermal heating (volume-weighted or mass-weighted) implemented in the with-cooling simulations, which is by default set to mass-weighted. In column~5, we denote the steady state rms Mach number ($\mathcal{M}_{\mathrm{hot}}$) of the hot phase gas ($T>10^7$~$\mathrm{K}$). In columns~6 and 7, we denote whether magnetic fields and cooling are switched on, respectively. Runs with magnetic fields are labelled by $\mathrm{mag}$ and runs with cooling are labelled with $\mathrm{f}$. Finally, in column~8, we show the end time of each simulation in $\mathrm{Gyr}$.\end{footnotesize} 
	
\end{table*}

\section{Results: turbulence without cooling}\label{sec:results-nocool}
In this section, we describe the results of our turbulence without cooling simulations. These control simulations are useful to understand the effects of magnetic fields and projection along the LOS on the structure functions in homogeneous isotropic turbulence. 
These will help us interpret our results from the more complex multiphase turbulence simulations.

\subsection{Velocity structure functions}\label{subsec:VSF}
According to the K41 model of homogeneous, isotropic, incompressible HD turbulence, the turbulent statistics in the inertial range, defined by the scales $L_{\mathrm{driv}} \gg \ell \gg \ell_\eta$ attain a universal form uniquely determined by the energy transfer rate ($\epsilon$). Here $\ell_\eta$ is the dissipation scale and $L_{\mathrm{driv}}$ is the driving scale of turbulence. Most of the scaling relations of K41 theory, such as slopes of power spectra and structure functions in turbulent statistics 
are valid in this range of scales. 

\begin{subequations}
For homogeneous and isotropic turbulence the longitudinal $\mathrm{VSF}$s depend only on the separation $r$. In this study, we define the first, second and $p^{\mathrm{th}}$ order $\mathrm{VSF}$s as \footnote{Note that the absolute value sign in \cref{eq:delv_VSF} is absent in fluid dynamics literature \citep{Pope2000}, but astrophysical studies (such as \citetalias{Li2020ApJ},\citetalias{Wang2021MNRAS}) use the definition of $\delta v_r$ in \cref{eq:delv_VSF}. But we have mainly studied the $\mathrm{VSF}_2$, where this difference is irrelevant. We also define the $\mathrm{VSF}_2$ as the trace of the second order velocity structure function tensor and do not decompose it into transverse and longitudinal components. However, we have verified that these components have similar scaling with $r$.} 
\begin{align}
&\mathrm{VSF}_1(r) = \langle \delta v_r \rangle,\label{eq:VSF_1}\\
&\mathrm{VSF}_2(r) = \langle \delta v_r^2 \rangle,\label{eq:VSF_2}\\ 
&\mathrm{VSF}_p(r) = \langle \delta v_r^p \rangle,\label{eq:VSF_p} \text{respectively, where}\\  
&\delta v_r = |\mathbf{v}(\mathbf{x}+\mathbf{e}_1r,t) - \mathbf{v}(\mathbf{x},t)|.\label{eq:delv_VSF}
\end{align}
$\mathbf{e}_1$ is a generalised unit vector and $\langle\rangle$ represents the ensemble average. 
In the inertial range, the $\mathrm{VSF}$s are expected to scale as: 
\begin{equation}
    \mathrm{VSF}_p =  C_p\epsilon^{p/3}r^{p/3},\label{eq:VSF_scaling}
\end{equation}
where $C_p$ is a constant. For $p$ different from 3, there are corrections to this scaling because of intermittency (\citealt{She1994}).
\end{subequations}

The first-order velocity structure function ($\mathrm{VSF}_1$) has been recently studied  in both observational \citepalias{Li2020ApJ} and numerical studies (\citealt{Hillel2020ApJ}; \citetalias{Wang2021MNRAS}). The second-order velocity structure function ($\mathrm{VSF}_2$) is simply related to the velocity correlation function
\begin{subequations}
\begin{equation}
\mathrm{VCF} (r) = \langle \mathbf{v}(\mathbf{x}+\mathbf{e}_1r,t) \cdot  \mathbf{v}(\mathbf{x},t) \rangle \label{eq:VCF}
\end{equation}
for isotropic homogeneous turbulence, for which the two are related as (e.g., see \citealt{Pope2000})
\begin{equation}
\mathrm{VCF} (r) = 2 v_\mathrm{rms}^2 - 2 \mathrm{VSF}_2(r),    
\end{equation}
where $v_\mathrm{rms}$ is the rms velocity. The velocity power spectrum is the Fourier transform of the velocity correlation function VCF, 
and describes the distribution of kinetic energy with scale. Therefore, we focus on 
the second order statistics for the remainder of the study but return to $\mathrm{VSF}_1$ in section \ref{subsec:comparison}. 
\end{subequations}

\subsubsection{Effect of resolution}\label{subsec:VSF_res_nocool}
\begin{figure}
		\centering
	\includegraphics[width=\columnwidth]{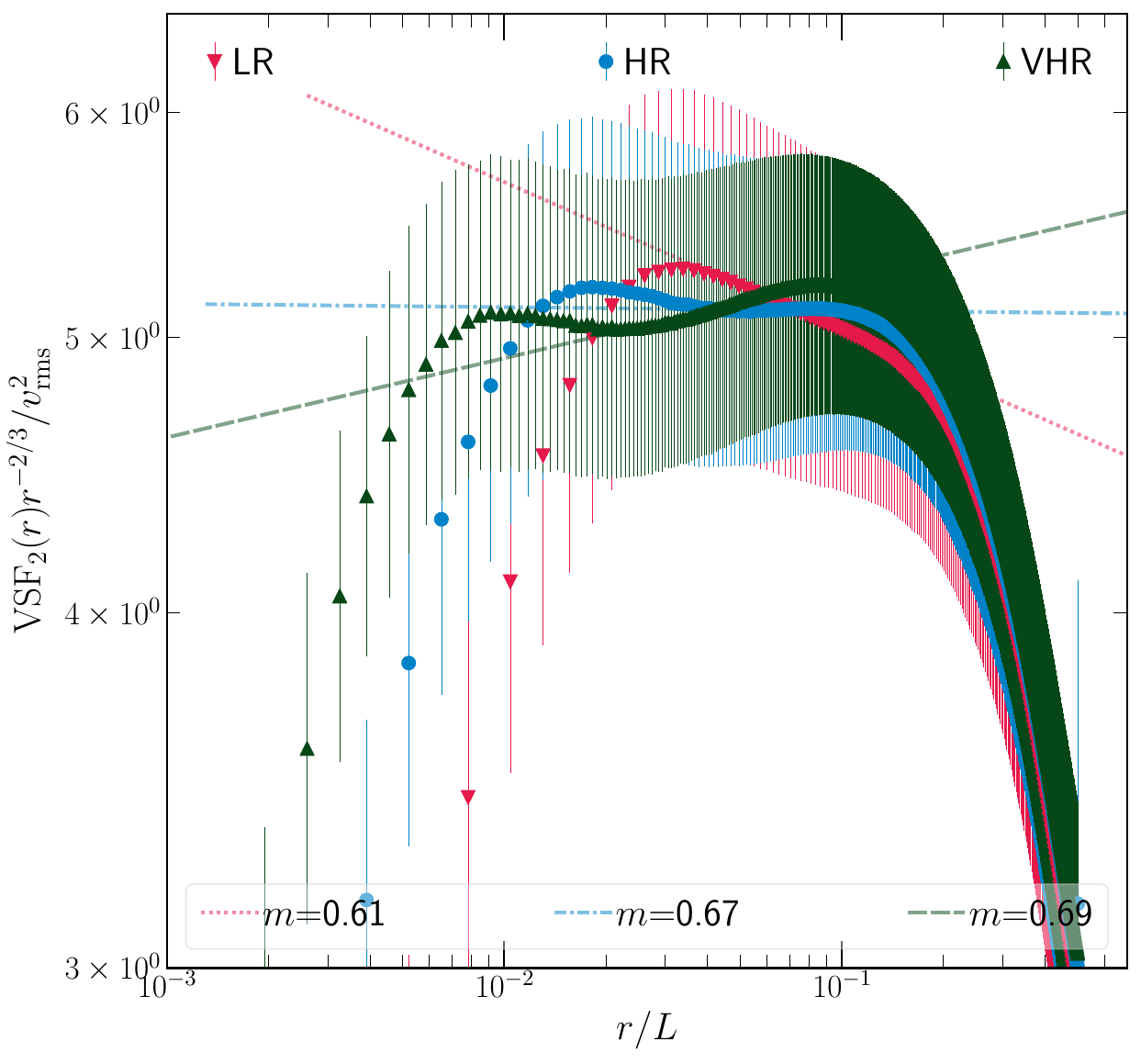}
	\caption[VSF for without cooling runs: Resolution]{The compensated and normalised second-order velocity structure function $(\mathrm{VSF_2})$ in the absence of cooling for three different resolutions LR, HR, VHR, averaged over five eddy turnover time scales. The errorbars indicate the temporal variation. The lines show the compensated fits to the $\mathrm{VSF}$s in the inertial range. The corresponding `m' values indicate the slopes of the uncompensated fits. The bottleneck effect dominates the inertial range in the LR run. The slopes of the $\mathrm{VSF}_2$ fits for the HR and VHR runs, are in close agreement with K41 theory, albeit somewhat steeper due to intermittency corrections \citep{She1994}. Note that the errorbars appear large since we show the compensated $\mathrm{VSF}_2$ with a small $y$ range in the plot.
	\label{fig:nocool_VSF_convergence}
	}
\end{figure}

Since we do not include explicit viscosity in our simulations, the viscous scale ($\ell_\eta$) is set by the spatial resolution in our simulations. The scale $\ell_\eta\propto\Delta x$, where $\Delta x$ is the size of a grid cell. The velocity power spectra and $\mathrm{VSF}$s flatten close to $\ell_\eta$ due to the bottleneck effect, which is seen in both numerical simulations and experiments 
\citep[although it is more prominent in the power spectra than in VSFs; e.g., ][]{schmidt2006numerical,Donzis2010JFM}. Hence in simulations we need to have enough spatial resolution to obtain convergent slopes of the $\mathrm{VSF}$s in the inertial range.

In \cref{fig:nocool_VSF_convergence}, we show the $\mathrm{VSF}_2$ compensated by the K41 slope ($r^{2/3}$) for three different resolutions. For these runs, the driving scale is the same, at $r/L=0.5$, whereas $\ell_\eta$ decreases with increasing resolution. The $\mathrm{VSF_{2}}$ flattens at the large scales, which corresponds to the driving scale of turbulence. At smaller scales, it shows a subtle rise, around $r/L \approx 0.02$ for the VHR run and $r/L \approx 0.04$ for the HR run. This rise is due to the bottleneck effect. In the LR run, the rise due to the bottleneck effect is close to the driving scale. Hence, the apparent slope of the $\mathrm{VSF}_2$ in the inertial range is shallower. 
The steep exponential dip in the $\mathrm{VSF_{2}}$ at small scales indicates the dissipation scale. 

We fit power-law functions to the $\mathrm{VSF_{2}}$, $0.04<r/L<0.1$ for the HR run and $0.02<r/L<0.1$ for the VHR run.
The convergent slope ($\sim0.69$) of the power-law fit is slightly steeper than the slope of the fit to our HR run, which is the standard resolution for all our simulations. The slight steepening from the K41 slope of $2/3$ in the VHR run may be due to the effects of intermittency \citep{She1994}. We use this resolution study to make sure that our turbulence with cooling runs from \citetalias{Mohapatra2021b} have enough separation of scales between $L_{\mathrm{driv}}$ and $\ell_\eta$.

\subsubsection{Effect of magnetic fields}\label{subsec:VSF_mag_nocool}
\begin{figure}
		\centering
	\includegraphics[width=\columnwidth]{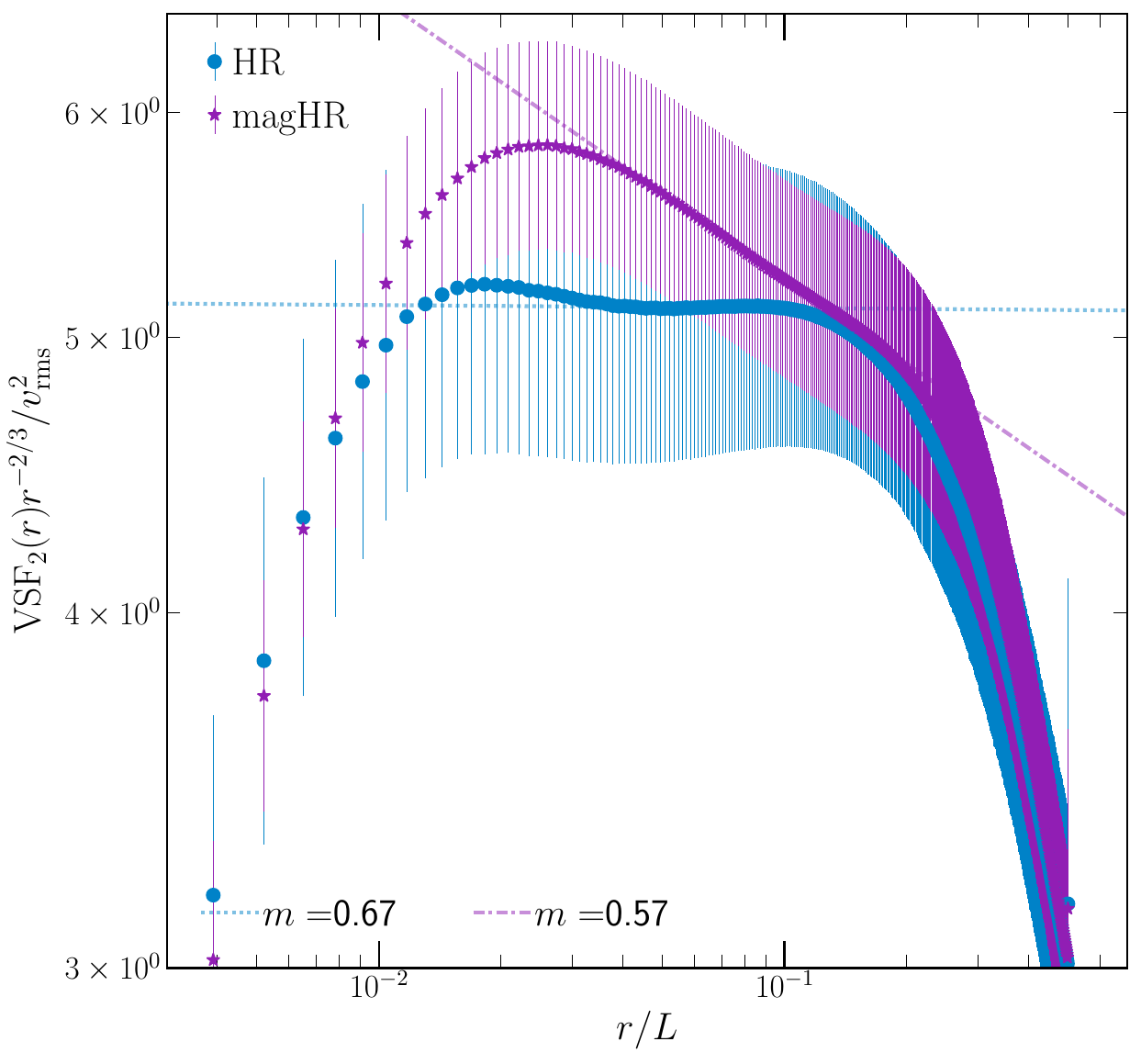}
	\caption[VSF of without cooling: magnetic field effects]{The compensated and normalised $\mathrm{VSF_2}$ of the turbulence without cooling runs with HD (HR) and MHD (magHR). The vertical bars indicate $1$-$\sigma$ error margins. We fit power-laws to the inertial range. The slopes to the uncompensated $\mathrm{VSF}$s are indicated as `m' values. The $\mathrm{VSF}_2$ is shallower in the MHD simulation and is closer to the IK scaling ($r^{0.5}$) than the GS scaling ($r^{2/3}$). Note that the errorbars appear large since we show the compensated $\mathrm{VSF_2}$ with a small $y$ range in the plot.
	\label{fig:nocool_VSF_MHD}
	}
\end{figure}

In \cref{fig:nocool_VSF_MHD}, we show the compensated $\mathrm{VSF}_2$ of the HR and magHR runs. The $\mathrm{VSF}_2$ becomes shallower for the magHR run. The slopes are slightly steeper than the scaling relation proposed by \cite{Iroshnikov1964SvA} and \cite{Kraichnan1965PhFl}, where $\mathrm{VSF}_2\propto\ell^{0.5}$ (also known as IK scaling). They are shallower than the scaling relation proposed by \cite{Goldreich1995ApJ}, where $\mathrm{VSF}_2\propto\ell^{2/3}$ (GS scaling).
The magHR run has small $\mathcal{M}\sim0.17$), rms Alfven Mach number (the ratio of rms velocity and rms Alfven velocity; $\mathcal{M}_A\sim1.56$) and $\beta\approx100$. In this parameter regime, MHD turbulence is nearly incompressible. The small scale eddies could be anisotropic, with different sizes perpendicular and parallel to the direction of the large-scale magnetic field. According to \cite{Boldyrev2005ApJ}, the scaling of the $\mathrm{VSF}_2$ is expected to lie in between the IK and GS scaling, and the exact exponent is determined by the anisotropy of turbulent fluctuations.  Similar to the HD runs, there maybe a slight steepening from the theoretical scaling due to the effect of turbulent intermittency, which has been proposed in \cite{Grauer1994PhLA,Politano1995PhRvE,Muller2000PhRvL}. More recently, \cite{Grete2020ApJ,Grete2021ApJ} using parameters similar to ours, have found the kinetic energy power spectrum to scale as $\propto k^{-4/3}$, which is also shallower than both $k^{-5/3}$ (GS) and $k^{-3/2}$ (IK) scaling\footnote{The M0.50P1.00isoA1.00H simulation from \cite{Grete2020ApJ} is comparable to our magHR run, but note that they use $\gamma=1$ and the magnetic field only has a mean component.}.

\subsubsection{Effect of projection}\label{subsec:VSF_proj_nocool}
\begin{figure}
		\centering
	\includegraphics[width=\columnwidth]{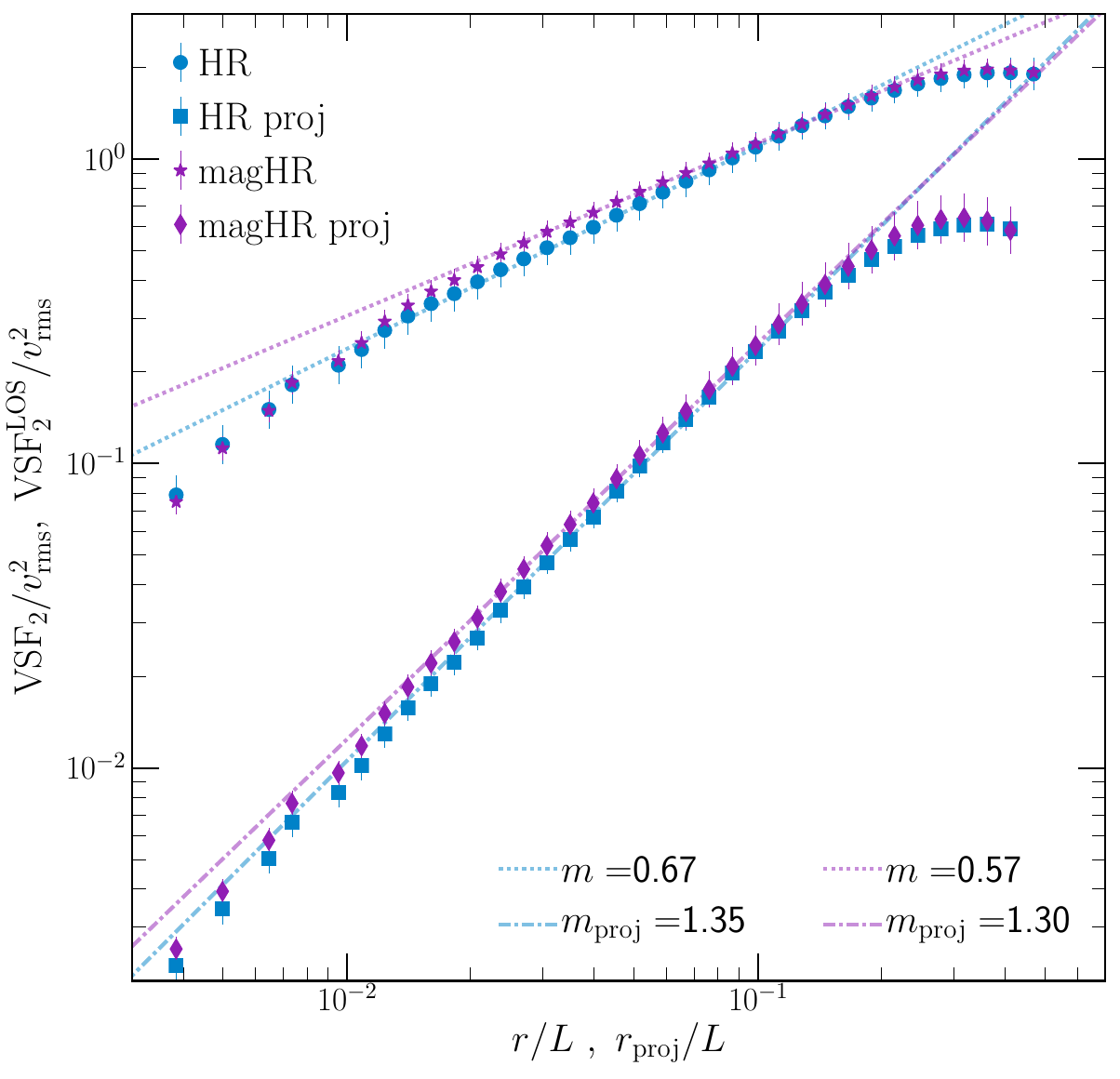}	
	\caption[VSF of without cooling runs:projection effect]{The $\mathrm{VSF}_2$ of the 3D velocity field (HR and magHR) and the projected 2D emission-weighted line of sight velocity fields ($\mathrm{VSF}_2^{\mathrm{LOS}}$, labelled HR proj and magHR proj) for turbulence without cooling runs, with and without magnetic fields. The lines indicate fits to the inertial range of the $\mathrm{VSF}$s. The projected $\mathrm{VSF}_2^{\mathrm{LOS}}$ along the $x-$, $y-$ and $z-$directions have been added. 
	The slopes of these fits are indicated by `$m$'. Upon projection, the $\mathrm{VSF}$s decrease in amplitude and become steeper, approximately by a factor of $r^{0.7}$ for both the runs. 
	\label{fig:nocool_VSF_proj}
	}
\end{figure}
In this section, we study the effect of projection on the $\mathrm{VSF}_2$, since we can only directly measure the projected LOS velocities through observations. Projection can generally have two different effects -- cancellation of velocities along the LOS, and smaller apparent projected distances compared to the actual 3D distances. The first removes power from small scales, since smaller eddies (being more numerous) are likely to have more cancellation along the LOS and this leads to the steepening of the $\mathrm{VSF}$. On the other hand, the smaller apparent projected distance shifts power from large scales to small scales, which can lead to flattening of the $\mathrm{VSF}$. 

In \cref{fig:nocool_VSF_proj}, we show the $\mathrm{VSF}_2$ of 
the 3D velocity field (HR and magHR) and the projected LOS velocity ($\mathrm{VSF}_2^{\mathrm{LOS}}$), normalised by the square of the 3D rms velocity $v_{\mathrm{rms}}^2$. The LOS velocity ($v_{\mathrm{LOS}}$) is weighted by emission (using to $Z_\odot/3$ emission table from \cite{Sutherland1993}). For example, $v_x$ projected along the $x$ direction is constructed as follows:
\begin{equation}
    v_{x,\mathrm{emw-proj-x}}(y,z)=\frac{\int_{-L/2}^{L/2} v_x\mathcal{L}\mathrm{d}x}{\int_{-L/2}^{L/2}\mathcal{L}\mathrm{d}x},\label{eq:emw-proj-vlos}
\end{equation}
where $\mathcal{L}=n_en_i\Lambda(T)$, $n_e$ and $n_i$ are electron and ion densities, respectively and $\Lambda(T)$ is the cooling function\footnote{Note that although we use an emission weight during projection, the runs in \cref{sec:results-nocool} do not have any radiative cooling, with the energy equation given by \cref{eq:energy_nocool}. Runs in \cref{sec:results-cool} have radiative cooling, with the energy equation given by \cref{eq:energy_cool}.} from \cite{Sutherland1993}.
We project the velocity along all three directions and 
sum their second order structure functions to improve statistics. 
The amplitude of $\mathrm{VSF}_2^{\mathrm{LOS}}$ close to the driving scale ($r/L\sim0.5$) is slightly smaller than $\mathrm{VSF}_2$. At scales smaller than the driving scale, the $\mathrm{VSF}_2^{\mathrm{LOS}}$ is also steeper than $\mathrm{VSF}_2$, roughly by a factor of $r^{0.7}$. This implies that cancellation along the LOS is the dominant effect. \cite{Zuhone2016APJ} derive a scaling relation for the projected $\mathrm{VSF}_2^{\mathrm{LOS}}$, and expect them to be steeper by a factor of $r$ compared to the 3D $\mathrm{VSF}_2$. The slopes of our $\mathrm{VSF}_2^{\mathrm{LOS}}$ are shallower than their predictions\footnote{We have also tested different projection weights, such as mass and volume, by replacing $\mathcal{L}$ with $\rho$ and unity in \cref{eq:emw-proj-vlos}, respectively, and find the steepening of the slope of the $\mathrm{VSF}_2^{\mathrm{LOS}}$ to be by $r^{0.7}$.}. \cite{Esquivel2005ApJ} use a 3D-Gaussian cube with a  3D spectral index of $-11/3$ (which corresponds to K41 spectral density). Upon projection, they also find the projected structure function to be steeper than the 3D structure function, but the steepening is less than the factor of $r$ (see their fig.~3).

\section{Results: turbulence with cooling}\label{sec:results-cool}
In this section we analyse the high-resolution turbulence with cooling runs from \citetalias{Mohapatra2021b}. In these simulations, we have implemented both the radiative cooling of the ICM and an idealised feedback heating loop, where the energy lost due to cooling is injected back into the gas in the form of turbulent kinetic energy and thermal heat. We implement two different models of thermal heat feedback, mass-weighted feedback ($Q_{\mathrm{mw}}$) and  volume-weighted feedback ($Q_{\mathrm{vw}}$). The details of our feedback loop implementation and our model setup are outlined in section 2.2.2 of \citetalias{Mohapatra2021b}. The energy equation for these runs is thus given by:
\begin{equation}
    \frac{\partial E}{\partial t}+\nabla\cdot ((E+P^*)\mathbf{v}-(\mathbf{B}\cdot\mathbf{v})\mathbf{B})=\rho\mathbf{F}\cdot\mathbf{v}+Q-\mathcal{L},\label{eq:energy_cool}
\end{equation}
where $Q$ is the thermal heating rate density and $\mathcal{L}$ is the radiative cooling rate density. The cooling rate $\mathcal{L}=0$ for $T<10^4$~$\mathrm{K}$.

For this part of the study, we have analysed three simulations with $f_{\mathrm{turb}}=0.1$, which is the ratio between the turbulent energy injection rate and the radiative cooling rate. The simulation parameters for these runs are listed in \cref{tab:sim_params}. The volume-weighted probability distribution functions (PDFs) of the mach number, temperature and pressure for these runs are shown in figure~12 of \citetalias{Mohapatra2021b}. In this study, we focus on the turbulent statistics of the velocity field of these simulations. 

\begin{figure*}
		\centering
	\includegraphics[width=2.0\columnwidth]{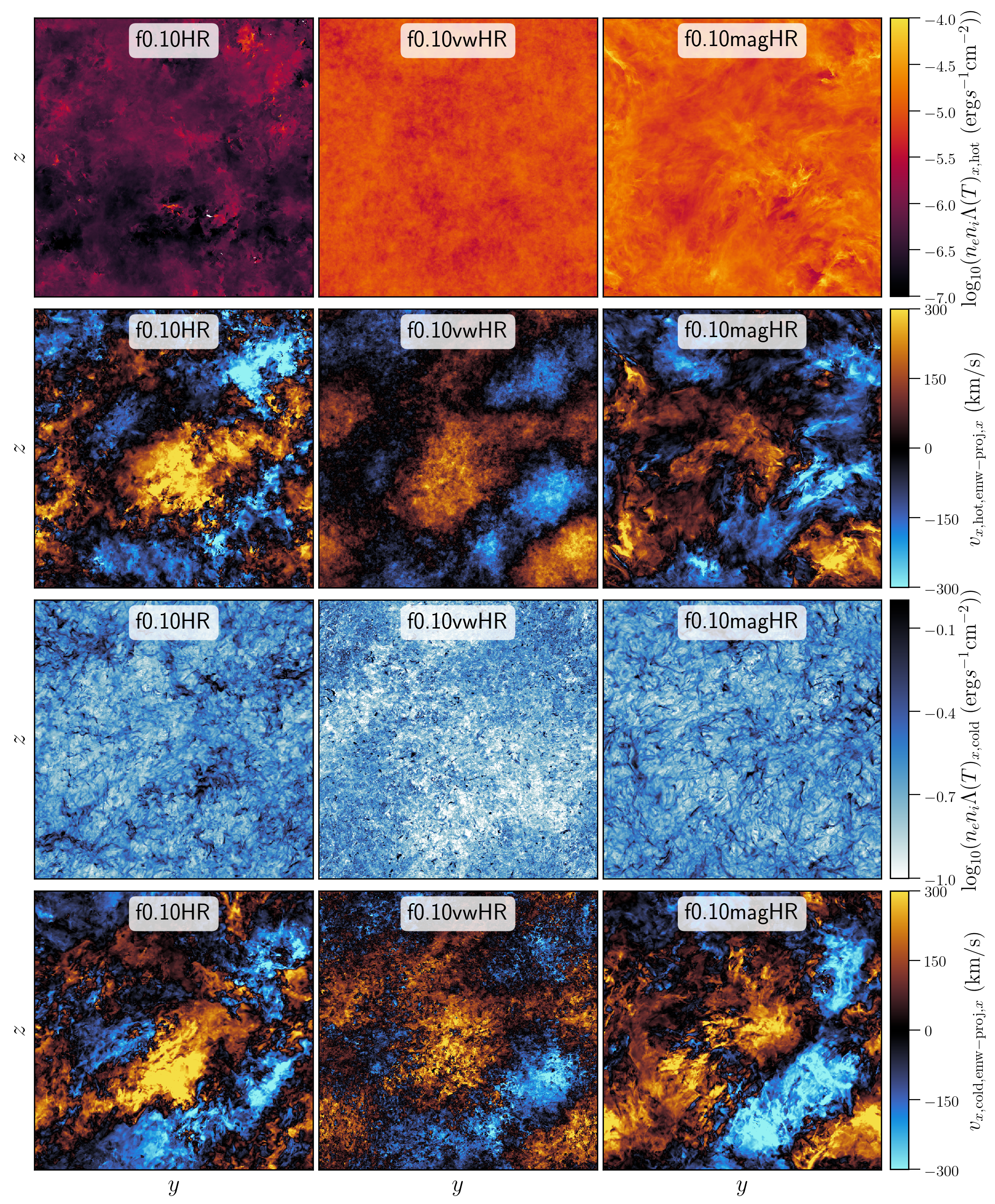}	
	\caption[Projection of emission weighted velocity plots]{First row: integrated total emission along the line of sight (here the $x$ direction) for gas in the hot phase ($T>10^7$~$\mathrm{K}$) for our turbulence with cooling runs: f0.10HR, f0.10vwHR, f0.10magHR at $t=0.911$~$\mathrm{Gyr}$. Second row: emission weighted line of sight velocity $v_{\mathrm{LOS}}$ projection maps ($v_{x,\mathrm{hot},\mathrm{emw-proj},x}$) for the hot phase at the same time snapshot.
	Third row: similar to first row, but showing emission from the cold phase gas ($10^4$~$\mathrm{K}<T<2\times10^4$~$\mathrm{K}$). Fourth row: similar to second row, but showing $v_{\mathrm{LOS}}$ of gas in the cold phase ($v_{x,\mathrm{cold},\mathrm{emw-proj},x}$). The colourbars at the right of each row represent the 
	surface brightness and the LOS velocities along the $x$ direction, respectively. We notice more small-scale features in all the projection plots of the f0.10vwHR run. The weak emission from the hot phase of f0.10HR is due to the lack of 
	gas above $10^7$~$\mathrm{K}$.
	\label{fig:velocity_projection_maps}
	}
\end{figure*}

\subsection{2D projection maps}\label{subsec:2D_proj}
In \cref{fig:velocity_projection_maps}, we show the projections of the mock X-ray emission from the hot gas and the net emission from the cooler H$\alpha$ emitting gas for our three runs at  $t=0.911$~$\mathrm{Gyr}$. We also present maps of the LOS velocity of gas emitting in these two wavelength ranges. We describe each of the rows of \cref{fig:velocity_projection_maps} in the following paragraphs. 

\subsubsection{Mock X-ray emission}\label{subsubsec:mock_xray_proj}
In the first row of \cref{fig:velocity_projection_maps}, we show the logarithm of LOS ($x$ direction) integrated total emission from gas in the X-ray emitting hot phase ($T>10^7$~$\mathrm{K}$). The net hot phase emission corresponds to the X-ray surface brightness measurements in \cite{churazov2012x,zhuravleva2014turbulent,zhuravleva2018}.

Among the different runs in our study, we observe X-ray emission among all sightlines for all three of our runs. But the net emission from the f0.10HR run is smaller than the rest, since it has less gas in the hot phase above the cutoff temperature of $10^7$~$\mathrm{K}$ (refer to fig.~12 of \citetalias{Mohapatra2021b}). Even though the three runs have similar amplitude of density fluctuations in the hot phase (fig.~7 of \citetalias{Mohapatra2021b}), the fluctuations in emission (approximately $\propto\rho^2\sqrt{T}$ for Bremsstrahlung emission from the hot phase) seem to be the largest for f0.10HR run and the smallest for the f0.10vwHR run. This could be attributed to the differing nature of fluctuations in the hot phase. In section 3.4 of \citetalias{Mohapatra2021b}, we showed that runs with small $f_{\mathrm{turb}}$ and $\mathcal{M}_{\mathrm{hot}}\ll1$ show isobaric density fluctuations in the hot phase, whereas runs with larger values $f_{\mathrm{turb}}$ and $\mathcal{M}_{\mathrm{hot}}\gtrsim1$ show adiabatic density fluctuations in the hot phase. For the f0.10HR run with transonic $\mathcal{M}_{\mathrm{hot}}=0.91$, the density fluctuations are expected to be more adiabatic, whereas for the f0.10vwHR run with subsonic $\mathcal{M}_{\mathrm{hot}}=0.23$, they are expected to be more isobaric. Adiabatic density fluctuations are expected to produce larger fluctuations in emission ($\propto \frac{\gamma+3}{2}\delta\rho/\rho$) compared to isobaric ones of the same amplitude ($\propto\frac{3}{2}\delta\rho/\rho$). The f0.10magHR run has $\mathcal{M}_{\mathrm{hot}}=0.59$ and shows trends in between the two extremes. 
The variations in emission are weak and at small scales in f0.10vwHR and occur at a larger scale for f0.10magHR and f0.10HR. We attempt to explain this feature later in this subsection.

\subsubsection{LOS velocity maps - hot phase}\label{subsubsec:vlos_hot}
In the second row of \cref{fig:velocity_projection_maps}, we show the projected LOS velocity of the hot phase gas, weighted by emission ($v_{x,\mathrm{hot},\mathrm{emw-proj},x}$). We construct this quantity by setting $\mathcal{L}=0$ for $T<10^7$~$\mathrm{K}$ in \cref{eq:emw-proj-vlos}. The LOS velocity of the hot phase could be measured by future high-resolution X-ray spectrographs such as XRISM\footnote{https://heasarc.gsfc.nasa.gov/docs/xrism} (and  \citealt{hitomi2016} in the recent past). These can give us useful information about turbulent velocities in the hot phase and their variation with scale.

Comparing these projected velocity maps among our simulations, 
we observe that their amplitude is largest for the f0.10HR run, smaller for f0.10magHR and smallest for the f0.10vwHR run. Similar to the hot phase mock X-ray emission maps, we find more small-scale structures in the f0.10vwHR run. The difference in the amplitudes of the emission weighted velocities  can be attributed to cancellation along the LOS.
The hot phase emission is mostly uniform and volume-filling at large scales for the f0.10vwHR run. But for the f0.10HR run, the hot-phase emission has non-uniform spatial distribution and peaks strongly in localised regions\footnote{Changing the hot-phase cutoff temperature to $10^6$~$\mathrm{K}$ results in the hot-phase emission showing stronger small-scale peaks, which are co-spatial with the peaks in the current emission plot (first row in \cref{fig:velocity_projection_maps}). This happens because the gas between $10^6$--$10^7$~$\mathrm{K}$ is  not volume-filling and is expected to occupy the transition layer between the hot and cold phases.}.  

\subsubsection{Mock optical emission}\label{subsubsec:mock_optical_proj}
In the third row of \cref{fig:velocity_projection_maps}, we show the  logarithm of the LOS integrated total emission from gas in the H$\alpha$ emitting cold phase ($10^4$~$\mathrm{K}<T<2\times10^4$~$\mathrm{K}$). This mock cold phase emission is comparable to the surface brightness of atomic filaments, which have been studied in many recent observations \citep{Gendron-Marsolais2018MNRAS,Olivares2019A&A,Boselli2019A&A}. 

Analysing the cold phase emission in our simulations, we find that the emission from the cold phase gas has a large area filling fraction for all runs. The emission peaks are filamentary for the f0.10HR and f0.10magHR runs, which have similar morphology as the H$\alpha$ filaments in the Perseus cluster \citep{Gendron-Marsolais2018MNRAS}. But for the f0.10vwHR run, the emission mainly comes from many smaller point sources. This cold-phase emission is more comparable to the H$\alpha$ surface brightness in Abell~2597 \citep{Tremblay2018ApJ} and Virgo \citep{Boselli2019A&A}, where smaller individual sources may be unresolved at the current angular resolution of these observations. The size of these cool clouds corresponds to the density of gas with $10^4$~$\mathrm{K}<T<2\times10^4$~$\mathrm{K}$ in these runs (see the density PDFs of these simulations in \cref{fig:dens_volume_PDF}). The f0.10vwHR run has denser gas in this temperature range, which correspond to smaller clouds. In comparison, the f0.10HR and f0.10magHR runs have rarer low temperature gas, which corresponds to more 
fluffy cool clouds.

\subsubsection{LOS velocity maps - cold phase}\label{subsubsec:vlos_cold}

In the fourth row of \cref{fig:velocity_projection_maps}, we show the projected LOS velocity of the cold phase gas, weighted by emission  ($v_{x,\mathrm{cold},\mathrm{emw-proj},x}$). We construct this quantity by setting $\mathcal{L}=0$ for $T>2\times10^4$~$\mathrm{K}$ and $T<10^4$~$\mathrm{K}$ in \cref{eq:emw-proj-vlos}. These correspond to the LOS velocity maps obtained from integral field unit spectroscopy (IFU) maps of the cold phase gas in observations such as \cite{Gendron-Marsolais2018MNRAS,Sarzi2018MNRAS,Tremblay2018ApJ,Boselli2019A&A} which were studied in \citetalias{Li2020ApJ}.

In our runs, we observe similar trends in the LOS velocity maps of hot and cold phases. The LOS velocities are smaller for the f0.10vwHR run compared to the other two runs. The small dense clouds in the f0.10vwHR run correspond to the small-scale features in the emission-weighted LOS velocities. These small-scale features are also seen in the emission and LOS velocities of the hot phase gas, which implies that cooling affects the morphology and kinematics of the hot phase 
as well, especially at small scales.

\subsection{Cold and hot gas velocity PDFs} \label{subsubsec:vel_pdf_cold_hot}
\begin{figure*}
    \centering
    \includegraphics[width=2.\columnwidth]{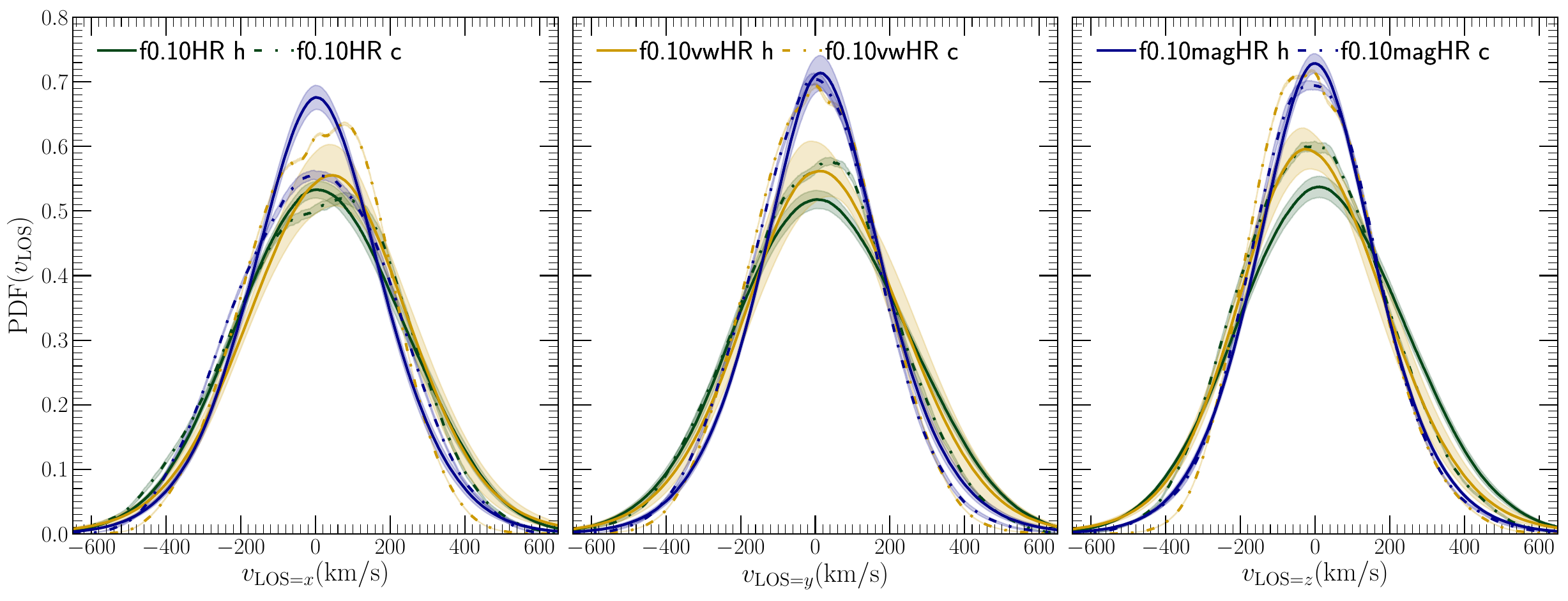}
    \caption[LOS Velocity pdf]{The volume PDF of LOS velocity for gas in the hot (solid lines) and cold phases (dash-dotted lines) for our three runs, averaged over 28 snapshots from $t=0.651$~$\mathrm{Gyr}$ to $t=1.003$~$\mathrm{Gyr}$. The three panels show the PDFs along the three directions ($x$, $y$ and $z$). The shaded regions around the PDFs denote the temporal variation in the PDF.  The hot phase velocities follow a normal distribution. The cold phase PDFs are flatter or distorted near the peak, which could correspond to the  
    bulk motion of the cold clouds and not their dispersion velocities, which could be much smaller. The rms velocity for the hot and cold phases for different runs are as follows:  f0.10HR $h:221\pm 15$~$\mathrm{km/s},\ c:205\pm 18$~$\mathrm{km/s}$; f0.10vwHR   $h:218\pm 12$~$\mathrm{km/s},\ c:167\pm 11$~$\mathrm{km/s}$;  f0.10magHR $h:185\pm 14$~$\mathrm{km/s},\ c:183\pm 20$~$\mathrm{km/s}$.}
    \label{fig:velocity_pdf}
\end{figure*}

In this subsection, we discuss the distribution of gas velocities in the hot and cold phases in our simulations. In \cref{fig:velocity_pdf}, we show the volume PDFs of these quantities for the three different directions. The solid lines represent hot gas PDFs and the dash-dotted lines represent the cold gas PDFs. The velocity dispersions of the gas in different phases are denoted in the caption of \cref{fig:velocity_pdf}. 

All the hot phase velocities show a smooth normal distribution. The widths of these PDFs are similar for all three directions for both hydro runs, which indicates isotropy. 
The cold phase velocity PDFs are somewhat distorted near the peaks. This could be due to the large-scale 
bulk motion of the clouds, which dominate the velocity dispersion in the PDFs. One can interpret them as a sum of many off-centre narrower Gaussians.

The difference between the widths of the PDFs is correlated to the ratio of densities of the hot and cold phases $\chi=\rho_{\mathrm{cold}}/\rho_{\mathrm{hot}}$ (see density PDF in \cref{fig:dens_volume_PDF}). 
We observe that the f0.10magHR run has equal velocity dispersion for both the hot and cold phases.  Since the amplitude of velocity dispersion is dominated by the contribution of large-scale modes, we infer that the hot and cold phases are coupled at large scales for our MHD run, which also has the smallest $\chi$. The smaller amplitude of velocity dispersion in this run 
is due to the conversion of turbulent kinetic energy into magnetic energy.

For the f0.10HR and f0.10vwHR runs, the cold phase has smaller velocity dispersion compared to the hot phase, implying that the cold phase has slower large-scale motion compared to the hot phase in these runs. The difference in the velocity dispersion of hot and cold phases is larger for the f0.10vwHR run, which also has a larger $\chi$.

\subsection{Velocity structure functions of multiphase gas } \label{subsec:vel_strufu_multiphase}
In this subsection, we study the $\mathrm{VSF}$s of the hot and cold phases in our simulations, which are the main focus of our study. We compare between the $\mathrm{VSF}$s of the two phases and their variation across our different simulations. We also study the effect of projection along the LOS on the $\mathrm{VSF}$s.

\subsubsection{Hot and cold gas velocity structure functions}\label{subsubsec:vel_strufu_cold_hot}
\begin{figure*}
		\centering
	\includegraphics[width=2.0\columnwidth]{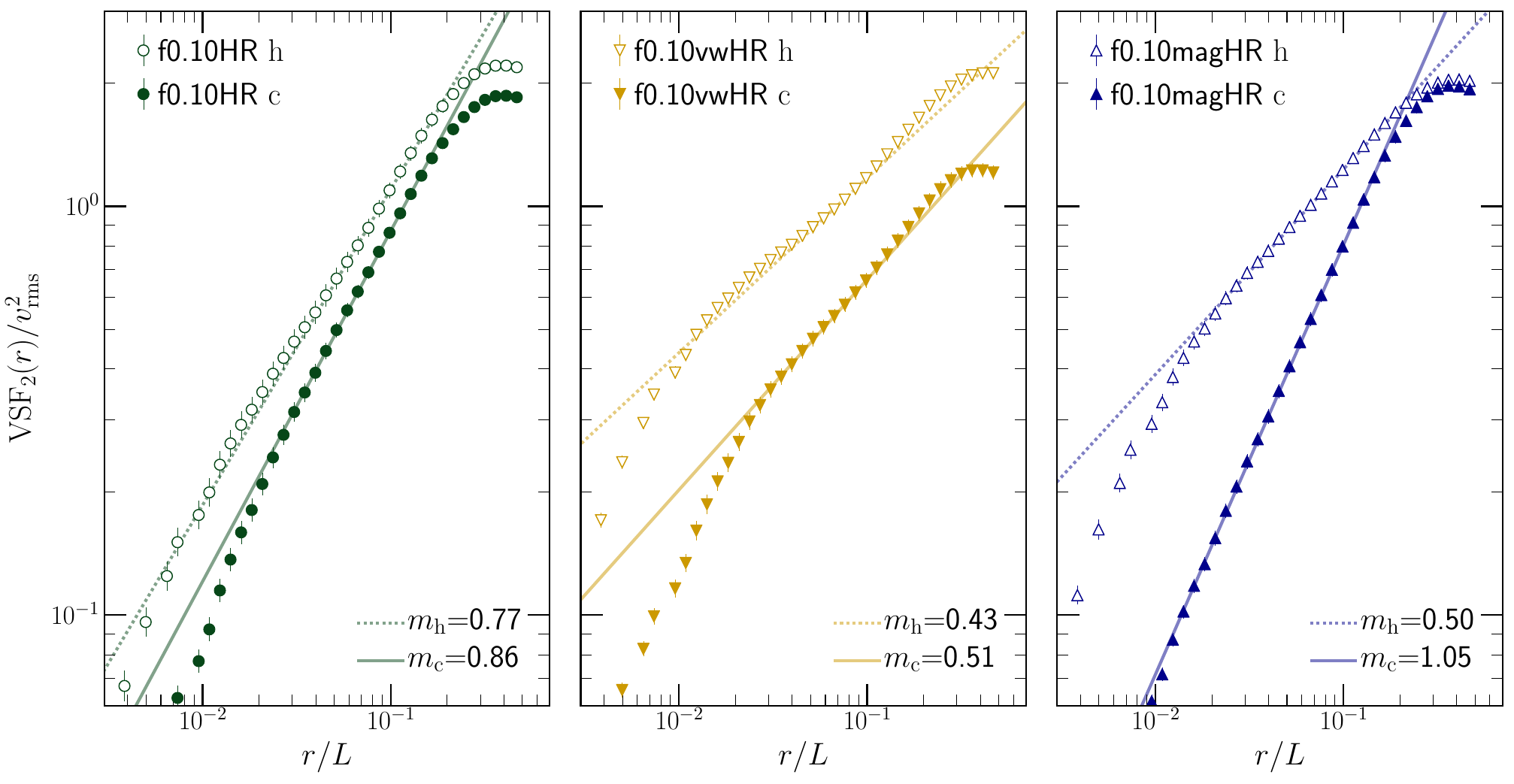}	
	\caption[VSF 3D hot and cold]{The $\mathrm{VSF}_2$ of the hot ($T > 10^7 K$, unfilled data points) and cold ($T < 2 \times 10^4 K$, filled data points) phase gas, in our three turbulence with cooling runs. The $\mathrm{VSF}_2$ are normalised by $v_{\mathrm{rms}}^2$. The error bars denote the temporal variation in the $\mathrm{VSF}_2$. The subscripts $\mathrm{h}$ (hot) and $\mathrm{c}$ (cold) in the legends denote the phase of the gas. We fit separate power-law scaling relations to the $\mathrm{VSF}_2$ of the two phases. We denote the scaling exponents by $m$ and the phases by the subscripts $h$ and $c$, and list them in the bottom right part of each panel. For the hydro runs, the $\mathrm{VSF}_2$ of the cold phase has a smaller amplitude than the $\mathrm{VSF}_2$ of the hot phase and is slightly steeper. But for the MHD run, the $\mathrm{VSF}_2$ of the cold phase has the same amplitude at large scales but follows a much steeper scaling relation at smaller scales. We discuss the possible implications of these findings in the main text. 
	
	\label{fig:vsf2_hot_cold_3d}
	}
\end{figure*}

\Cref{fig:vsf2_hot_cold_3d} shows $\mathrm{VSF}_2$ of the hot ($T > 10^7 K$) and cold ($T < 2 \times 10^4 K$) phase gas in the three different with cooling runs. The $\mathrm{VSF}_2$ are normalised by the square of the rms velocity of all gas ($v_{\mathrm{rms}}^2$). The unfilled data points represent the hot phase $\mathrm{VSF}_2$ and the filled data points represent the cold phase $\mathrm{VSF}_2$. 
Similar to the turbulence without cooling runs in \cref{sec:results-nocool}, the $\mathrm{VSF}_2$ for both hot and cold phases increases with increasing $r$ and flattens close to the driving scale ($r/L\approx0.5$).

We fit separate power-law functions to the scaling range of the $\mathrm{VSF}_2$ for the hot and cold phases. We have denoted the exponents of these functions as $m_h$ and $m_c$ in the bottom right side of the figure panels for the fits to the hot and cold phase $\mathrm{VSF}_2$, respectively. We have also listed these scaling exponents in \cref{tab:VSF_fit_params}. 

\paragraph{f0.10HR}\label{par:vsf2_hc3d_f0.10HR}For the f0.10HR run, the hot phase $\mathrm{VSF}_2$ is steeper than the K41 scaling of $r^{2/3}$. It is also steeper than the turbulence without cooling run (HR) in \cref{subsec:VSF}. The steepening of the $\mathrm{VSF}_2$ could be due to transonic turbulence in the hot phase ($\mathcal{M}_{\mathrm{hot}}=0.91$), where the slopes of the $\mathrm{VSF}_2$ could be approaching the Burgers turbulence slopes \citep{Burgers1948171} of $r^1$, which is valid for compressible supersonic turbulence \citep{Federrath2013}. 
Our scaling exponents are consistent with the $\mathrm{VSF}_2$ slopes in the simulations of \cite{Schmidt2008} and the $\mathrm{VSF}_1$ slope in the transonic regime of \cite{Federrath2021NatAs} (see fig.~2 of their paper)\footnote{See \cref{fig:vsf1_coldp_hot3d_obs} for a direct comparison of the $\mathrm{VSF}_1$ slope with \cite{Federrath2021NatAs}.}.

The $\mathrm{VSF}_2$ for the cold phase has a smaller amplitude at the driving scale, consistent with the smaller velocity dispersion for the cold phase in \cref{fig:velocity_pdf}. At scales smaller than the driving scale, the cold phase $\mathrm{VSF}_2$ has a slightly steeper slope compared to the hot phase $\mathrm{VSF}_2$. 

\paragraph{f0.10vwHR}\label{par:vsf2_hc3d_f0.10vwHR} For the f0.10vwHR run, the hot phase $\mathrm{VSF}_2$ is much shallower than the K41 scaling.
Since this run has a small $\mathcal{M}_{\mathrm{hot}}=0.23$, we expect the velocity field in this run to have significant contribution from gas motions due to thermal instability. These velocities are generated when the gas separates into hot and cold phases due to thermal instability, even in the absence of external turbulent driving. From the f0.001 run (with almost no turbulent driving) in \citetalias{Mohapatra2021b}, we find the velocities due to thermal instability to be $\approx120~\mathrm{km/s}$. This approximately corresponds to the isobaric collapse of a region of size $\sim1~\mathrm{kpc}$ in a thermal instability time-scale ($t_{\mathrm{TI}}\approx57\mathrm{Myr}$; see eq.~6b in \citetalias{Mohapatra2021b}). The velocity generated by thermal instability depends on the size of the cooling region ($r_{\rm cool}$) and the cooling time of intermediate gas ($t_{\rm cool}$) via the dimensionless number $r_{\rm cool}/c_s t_{\rm cool}$, a larger value of which results in faster flows and higher amplitude fluctuations (\citealt{Dutta2021}). These velocities are supposed to have a flat $\mathrm{VSF}_2\propto r^0$ (with no structure across scales)\footnote{The velocities due to thermal instability are expected to have $\mathrm{VSF}_2\propto r^0$, since the growth rate of thermal instability is independent of scale ($t_{\mathrm{TI}}$ is defined in eq.~6b of \citetalias{Mohapatra2021b}) in the absence of thermal conduction. So both large and small scale isobaric modes grow at the same rate, as seen in fig.~2 of \cite{sharma2010thermal}.  \citet{Das2021MNRAS} consider isochoric modes, which grow more slowly, but are not relevant for the temperatures and length scales of cool cores.}. Such velocities increase the amplitude of small-scale velocity fluctuations, which flatten the $\mathrm{VSF}_2$.

The cold phase $\mathrm{VSF}_2$ has a much lower amplitude compared to the hot phase $\mathrm{VSF}_2$. This behaviour is expected since the value of $\chi$ is large, i.e., the cold phase is much denser than the hot phase. Although the amplitude of turbulent acceleration is the same for the two phases, the cold/dense phase with a small volume filling factor has a smaller turbulent velocity. Similar to the f0.10HR run, the cold phase $\mathrm{VSF}_2$ is slightly steeper than the hot phase $\mathrm{VSF}_2$. 

The trends in the $\mathrm{VSF}_2$ of the cold phase for both of our hydrodynamic runs (f0.10HR and f0.10vwHR) are similar to the results of \cite{Gronke2021arXiv} (see their fig.~19). Using a setup similar to ours, they show that the cold phase $\mathrm{VSF}_2$ has a smaller amplitude compared to the $\mathrm{VSF}_2$ of the rest of the gas but has similar scaling with $r$. 


\paragraph{f0.10magHR}\label{par:vsf2_hc3d_f0.10magHR} For the f0.10magHR run, the hot phase $\mathrm{VSF}_2$ scales with $r$ as $r^{0.5}$. The rms Alfv\'en Mach number of the hot phase $\mathcal{M}_{A,\mathrm{hot}}\sim0.92$. So the $\mathrm{VSF}_2$ of the hot phase is slightly flatter than the magHR run (without cooling MHD simulation in \cref{subsec:VSF_mag_nocool}) and is closer to the IK scaling. The hot phase $\mathrm{VSF}_2$ could also flatten due to the effect of thermal instability induced small-scale velocities, as we discussed in the paragraphs above.

The cold phase $\mathrm{VSF}_2$ has the same amplitude as the hot phase $\mathrm{VSF}_2$ at large scales. The large-scale velocities of the cold and hot phases are coupled due to magnetic fields. But at scales smaller than the driving scale, the cold phase $\mathrm{VSF}_2$ is much steeper than the hot phase $\mathrm{VSF}_2$ (more than twice as steep). 

The scaling of the $\mathrm{VSF}_2$ in this run can be attributed to the effect of magnetic fields. By Alfv\'en's theorem of flux freezing, the magnetic field lines are frozen into the gas as it condenses out of the hot phase into to the small-scale cold clouds. The amplification of the magnetic field due to flux freezing depends on $\chi$ as $\chi^{2/3}$. The cold phase is dominated by magnetic pressure, with $\beta\lesssim0.1$ (see fig.~13 of \citetalias{Mohapatra2021b}) and the gas is constrained to move along the magnetic field lines. This effect becomes stronger at small scales. It reduces the gas velocities at these scales and makes the $\mathrm{VSF}_2$ steep.

\subsubsection{Projected hot and cold gas velocity structure functions}\label{subsubsec:proj_vel_strufu_cold_hot}
\begin{figure*}
		\centering
	\includegraphics[width=2\columnwidth]{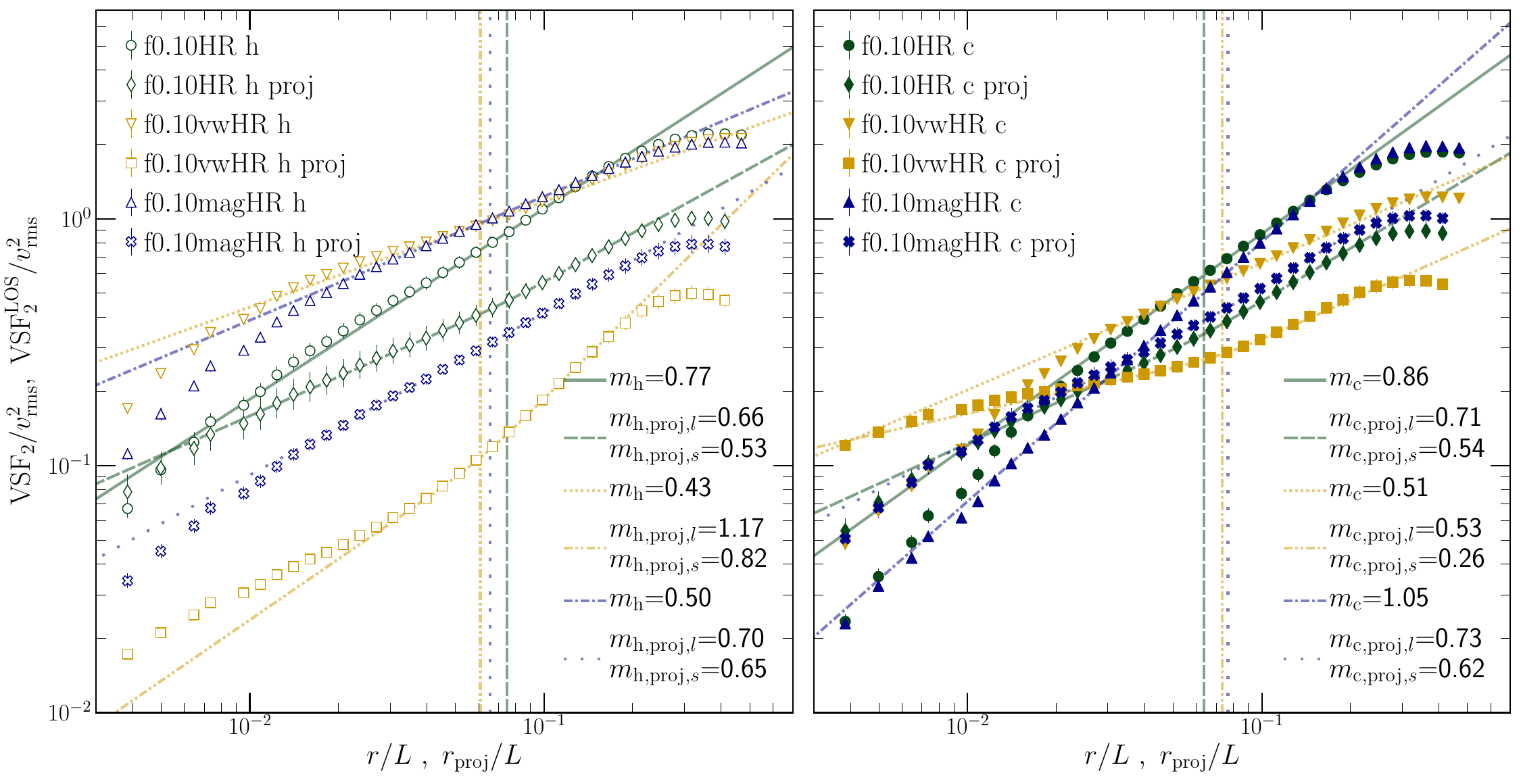}	
	\caption[VSF of with cooling runs: projection]{The second-order velocity structure function of  the 3D velocity field ($\mathrm{VSF}_2$) and the projected LOS velocity field ($\mathrm{VSF}_2^{\mathrm{LOS}}$, denoted by `proj' in the legend) for our turbulence with cooling runs. We have summed the $\mathrm{VSF}_2^{\mathrm{LOS}}$ along our three LOS directions $x$, $y$ and $z$. 
	In the left panel, we show these quantities for the hot phase gas and in the right panel for the cold phase gas.
	All the $\mathrm{VSF}_2$ are normalised by $v_{\mathrm{rms}}^2$. 
	We fit broken power-law functions to the $\mathrm{VSF}_2^{\mathrm{LOS}}$ of both hot and cold phases, and their slopes are denoted by $m_{\mathrm{proj}}$. The scaling exponents at large and small scales, are denoted by the subscripts $l$ and $s$, respectively and are shown in the bottom right of each panel. The break scales are shown as the vertical lines. We have also listed them in \cref{tab:VSF_fit_params}. 
	\emph{Left panel:}
	For the f0.10HR run, the $\mathrm{VSF}_2^{\mathrm{LOS}}$ is slightly flatter than the $\mathrm{VSF}_2$ for the hot phase gas. It steepens slightly for the f0.10magHR run and becomes significantly steeper for the f0.10vwHR run upon projection. 
	\emph{Right panel:}
	For the cold phase, the $\mathrm{VSF}_2^{\mathrm{LOS}}$ for the f0.10HR and f0.10magHR is shallower than the $\mathrm{VSF}_2$. But for the f0.10vwHR run, it is roughly parallel to the $\mathrm{VSF}_2$ at large scales and shallower at small scales. We discuss these findings in detail in the main text.
	\label{fig:vsf2_hc_projection}
	}
\end{figure*}

In this subsection, we show the effect of projection along the LOS on the $\mathrm{VSF}_2$ of the hot and cold phases. In \cref{fig:vsf2_hc_projection}, we show the $\mathrm{VSF}_2^{\mathrm{LOS}}$ and the $\mathrm{VSF}_2$ for the hot (left panel) and cold phases (right panel). 
The $\mathrm{VSF}_2^{\mathrm{LOS}}$ are constructed using emission-weighted LOS velocities, snapshots of which are shown in the second and fourth rows of \cref{fig:velocity_projection_maps}.
We have summed the $\mathrm{VSF}_2^{\mathrm{LOS}}$ along 
$x-$, $y-$ and $z-$ axes and normalised them by the 3D $v_{\mathrm{rms}}^2$, so that we can directly compare them with the 3D $\mathrm{VSF}_2$ and look for the effects of projection. We have also averaged them temporally over 28 snapshots from $t=0.651$~$\mathrm{Gyr}$ to $t=1.003$~$\mathrm{Gyr}$.  In all cases, the projected $\mathrm{VSF}_2^{\mathrm{LOS}}$ increase with increasing $r_{\mathrm{proj}}$ and peak close to the driving scale, where $r_{\mathrm{proj}}$ is the  separation between two points along the projection plane. The amplitude of the $\mathrm{VSF}_2^{\mathrm{LOS}}$ is slightly smaller than the $\mathrm{VSF}_2$ at the driving scale for all of our runs because power at large scales is shifted to smaller projected separations. In contrast to the $\mathrm{VSF}_2$, the $\mathrm{VSF}_2^{\mathrm{LOS}}$ flattens at small scales inside the scaling range and is 
better fit by a broken power-law function $\mathrm{SBP}(r_{\mathrm{proj}})$, which is described in \cref{eq:sbp_fit}. The flattening of the $\mathrm{VSF}_2^{\mathrm{LOS}}$ at small scales is due to insufficient cancellation of the small-scale eddies upon projection, especially for the cold phase. We discuss the effects of projection in detail for these two phases in the next two paragraphs.

\paragraph{Effect of projection on the hot phase VSFs}\label{par:proj_hot_phase} 
Here we discuss the $\mathrm{VSF}_2^{\mathrm{LOS}}$ for the hot phase gas. Although the amplitudes of the normalised $\mathrm{VSF}_2$ are approximately equal at the driving scale, the amplitudes of the normalised $\mathrm{VSF}_2^{\mathrm{LOS}}$ are smaller and vary for our three runs. These trends are similar to the variation in the amplitudes of the projected LOS hot phase velocity maps, shown in the second row of \cref{fig:velocity_projection_maps}. The hot phase gas in the f0.10HR run is clumpy and has strong localised emission features 
surrounding the cold/dense strongly emitting gas (see 
\cref{fig:velocity_projection_maps}). Observations of cool-core halos (e.g., \citealt{Fabian2003}, \citealt{Werner2014MNRAS}), multiphase galactic outflows (e.g., \citealt{Strickland2004}), and ram pressure-stripped tails of galaxies moving through the ICM (e.g., \citealt{Sun2021}) show similar spatial correlation between different temperature phases. Along any particular LOS, we expect to find only few such bright 
patches of hot phase gas mostly around cold clouds. Due to smaller number of eddies from these 
patches, the LOS velocities do not cancel out upon projection. In contrast, the hot phase gas in the f0.10vwHR run is more volume-filling and shows variations only at small scales (see middle columns of \cref{fig:velocity_projection_maps}). In this case, we have larger number of cancelling eddies during projection and as a result the  $\mathrm{VSF}_2^{\mathrm{LOS}}$ is steeper than the $\mathrm{VSF}_2$. 
For the f0.10magHR run, the emission features lie in between the two extremes, and it shows weak cancellation of LOS velocity upon projection. This leads to steeper $\mathrm{VSF}_2^{\mathrm{LOS}}$ than $\mathrm{VSF}_2$.

\paragraph{Effect of projection on the cold phase VSFs}\label{par:proj_cold_phase}
Here we discuss the $\mathrm{VSF}_2^{\mathrm{LOS}}$ and its relation with the $\mathrm{VSF}_2$ for gas in the cold phase. These two quantities are plotted in the right panel of \cref{fig:vsf2_hc_projection}. In all our runs, the $\mathrm{VSF}_2^{\mathrm{LOS}}$ becomes flatter than the $\mathrm{VSF}_2$ and has larger values at small scales. The flattening effect of projection is much stronger at small scales because the cold phase gas is distributed into small-sized clouds (see the third row of \cref{fig:velocity_projection_maps}). Since this phase is not volume-filling, we find few cold clouds along any particular LOS during projection.
Hence the cancellation of eddies upon projection along the LOS is small. In addition to that, the distance between any of these cold clumps in the projected plane is also smaller than the real 3D distance ($r_{\mathrm{proj}}\leq r$). This can make the $\mathrm{VSF}_2^{\mathrm{LOS}}$ larger at small $r_{\mathrm{proj}}$. These effects make the $\mathrm{VSF}_2^{\mathrm{LOS}}$ shallower than the $\mathrm{VSF}_2$, especially for the f0.10magHR run which has a steep 3D $\mathrm{VSF}_2$.
For the fiducial f0.10HR run, the $\mathrm{VSF}_2^{\mathrm{LOS}}$ is slightly flatter than the $\mathrm{VSF}_2$ at large scales. The f0.10vwHR run, which has the smallest cool clouds among our three runs, shows the flattest $\mathrm{VSF}_2^{\mathrm{LOS}}$ slopes at small scales. But at large scales, the $\mathrm{VSF}_2^{\mathrm{LOS}}$ is roughly parallel to the $\mathrm{VSF}_2$.

\section{Discussions}\label{sec:discussions}
In this work, we have two key aims: (i) to study the relation between the 3D $\mathrm{VSF}$s of hot and cold phases and (ii) to understand the effect of projection along the LOS on these $\mathrm{VSF}$s. Our findings are important to understand the relation between the $\mathrm{VSF}$s studied in observational studies (e.g.,  \citetalias{Li2020ApJ}) of the atomic gas and future X-ray observations of the hot phase gas (such as with XRISM) and the 3D velocity structure of the hot ICM gas. 

In the first part of this study, we studied the effect of magnetic fields and the effect of projection on the $\mathrm{VSF}$s in homogeneous idealised turbulence simulations without cooling. In the second part, we analysed three high resolution multiphase ICM simulations with global energy balance from \citetalias{Mohapatra2021b}. Here we discuss the effects of important physical elements such as cooling and magnetic fields, and our methods such as nature of thermal feedback energy on our results. We also discuss the effect of projection for both parts of our study and finally compare our results with other simulations and observations.

\subsection{Effect of cooling physics and thermal feedback models}\label{subsec:cooling_effects_mw_vw}
The $\mathrm{VSF}_2$ in part 1 of our study follow K41 scaling for the HD run and closer to IK scaling for the MHD run, which are according to theoretical expectations. In part 2 of this study, cooling and thermal instability clearly affect the the scaling and amplitude of the $\mathrm{VSF}_2$ in both hot and cold phases. 

In addition to separating the gas into hot and cold phases, thermal instability also introduces small velocities in the gas during phase separation. These velocities are expected to have a flat $\mathrm{VSF}_2$, similar to white noise, since thermal instability in absence of thermal conduction has the same growth rate at all scales. For turbulence at small and intermediate $\mathcal{M}_{\mathrm{hot}}$, such as in f0.10vwHR and f0.10magHR runs, this flattens the $\mathrm{VSF}_2$ of the hot phase. 

Our two different feedback models lead to different density distributions and different velocity dispersion in the hot and cold phases. The f0.10vwHR run has denser cold phase gas (larger $\chi$) than the f0.10HR run, whereas their hot phase density is similar. At scales larger than the size of the cool clouds, we expect the turbulent cascade to be established in the hot phase. Then at intermediate scales, the hot phase gas drags the cold clouds along with it. This imprints the velocity structure of the hot phase into the cold phase, making their $\mathrm{VSF}_2$ scale similar to the $\mathrm{VSF}_2$ of the hot phase. The small scale eddies maybe less efficient at dragging cold clouds of comparable size, which can explain why the cold phase $\mathrm{VSF}_2$ are slightly steeper than the hot phase $\mathrm{VSF}_2$. The effect of this drag acceleration, which should depend on scale, is weaker if the density contrast between the two phases is higher, as seen in the f0.10vwHR run. 
A similar effect is also seen in the simulations of \cite{Gronke2021arXiv}, where a larger density contrast between hot and cold phases leads to weaker mixing between them. \cite{Gronke2021arXiv} also find the cold phase $\mathrm{VSF}_2$ to be roughly parallel to the $\mathrm{VSF}_2$ of rest of the gas but slightly smaller in amplitude.

\subsubsection{Driving turbulence only in the hot phase}\label{subsubsec:hot_driv}
\begin{figure}
		\centering
	\includegraphics[width=\columnwidth]{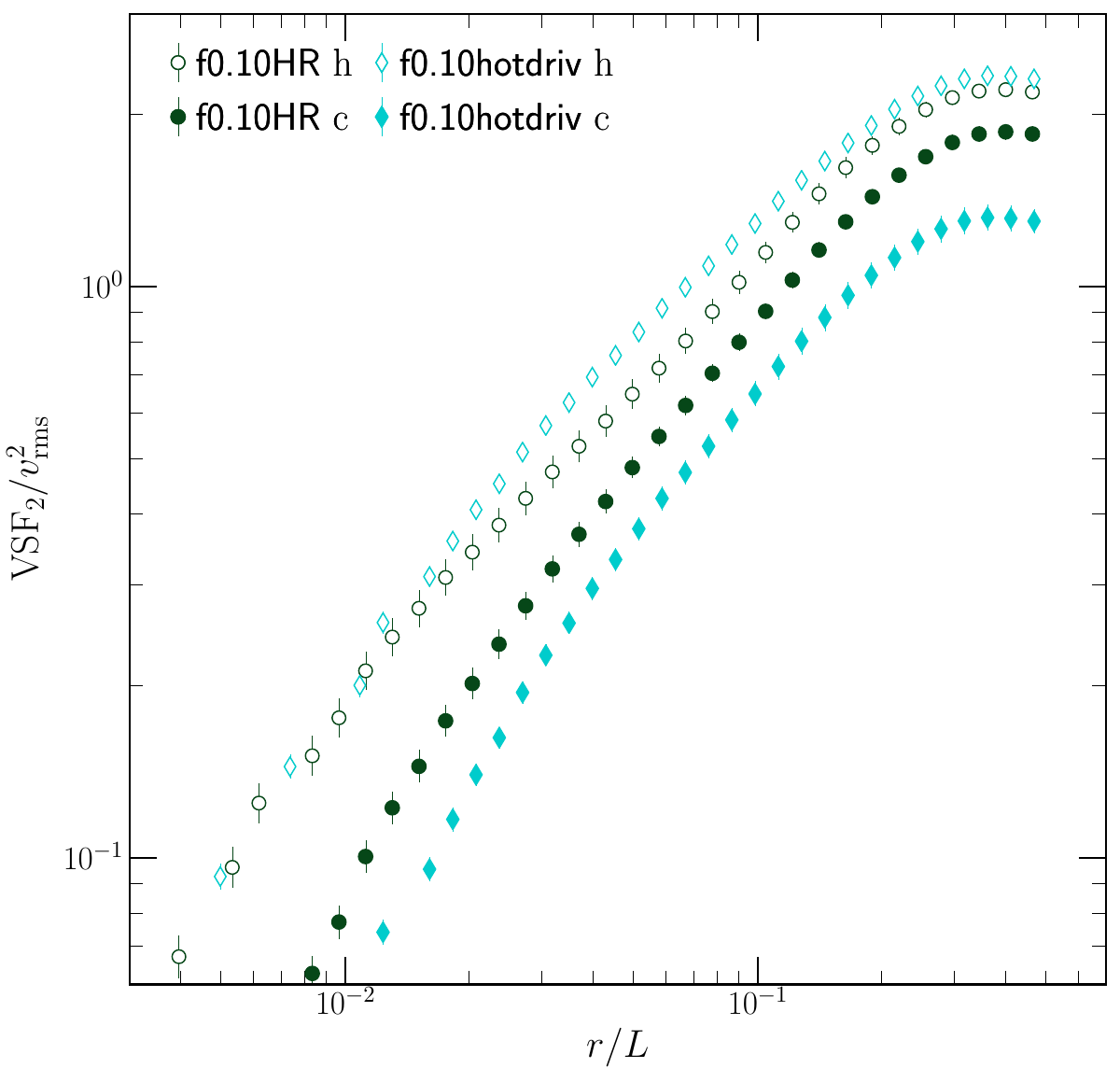}	
	\caption[VSF of hotdriv run]{Hot and cold phase $\mathrm{VSF}_2$ where we drive turbulence only in gas with $T>10^6$~$\mathrm{K}$, shown in cyan. We also show the fiducial f0.10HR run for comparison. The cold phase $\mathrm{VSF}_2$ is slightly smaller since it is not driven directly. The hot phase $\mathrm{VSF}_2$ is larger since we are only driving the hot phase, while maintaining global thermal balance with 10\% heating provided by turbulence. 
	The $\mathrm{VSF}_2$ of the two phases are still roughly parallel to each other.
	\label{fig:vsf2_hotdriv}
	}
\end{figure}

To understand the effect of this drag, we also conducted a lower resolution simulation ($384^3$ resolution elements) where we drive turbulence only in gas with $T>10^6$~$\mathrm{K}$, with all other conditions similar to f0.10HR run. We show the $\mathrm{VSF}_2$ for this run in \cref{fig:vsf2_hotdriv}. The cold phase $\mathrm{VSF}_2$ is still parallel to the hot phase $\mathrm{VSF}_2$ at all scales for this run, denoting the importance of hydrodynamic drag at all scales. The amplitude of the $\mathrm{VSF}_2$ of the cold phase gas is slightly smaller, mainly because we do not impart any velocities to the cold phase at the driving scale, and cold phase is driven only by drag from the hot phase.

\subsubsection{Spatial correlation function between cold and hot phase velocities}\label{subsubsec:cross_corr_cold_hot}

\begin{figure}
		\centering
	\includegraphics[width=\columnwidth]{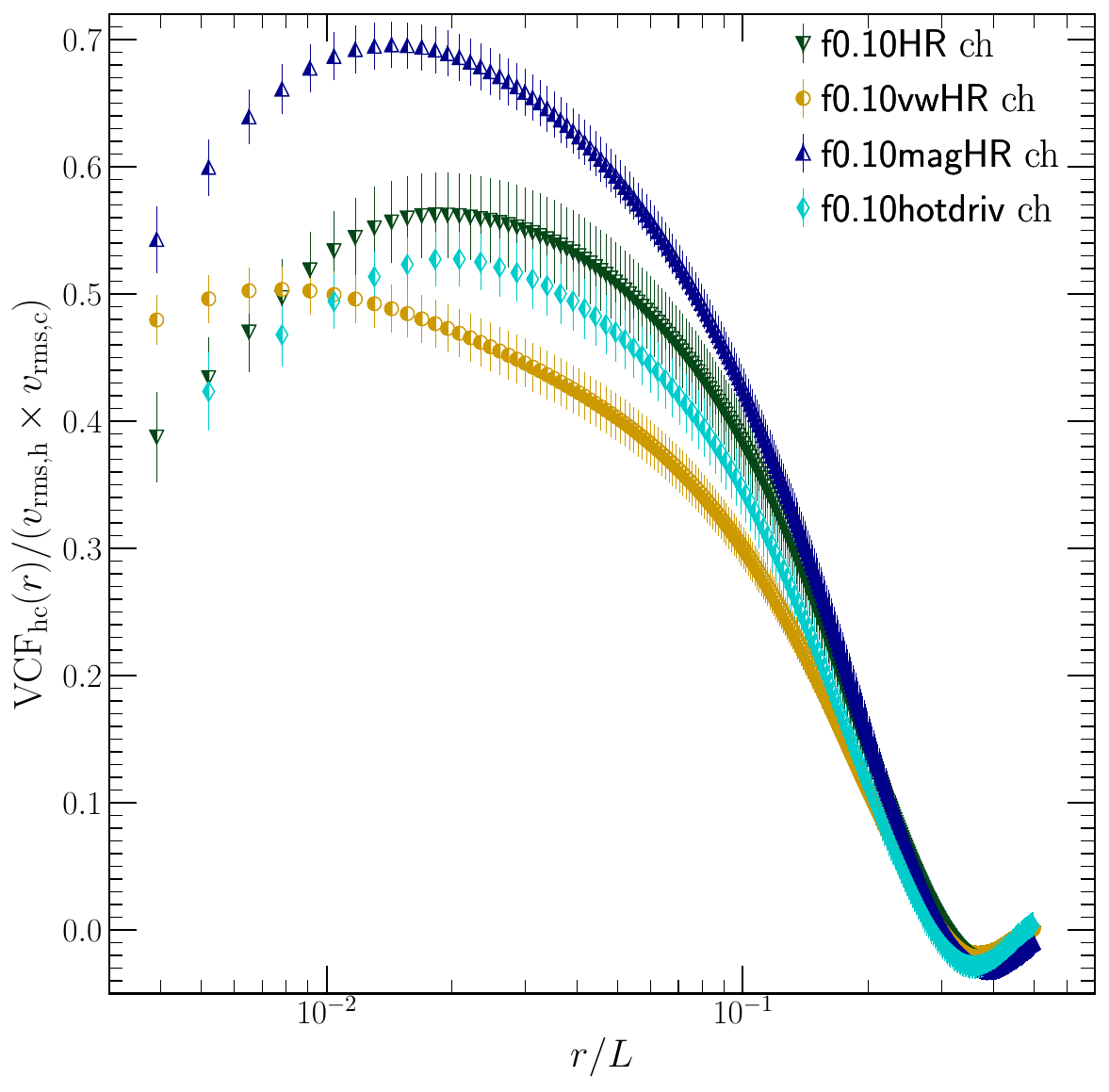}	
	\caption[VCF of cold and hot gas]{The two point cross correlation function $\mathrm{VCF}_{\mathrm{hc}}(r)$ (see equation \ref{eq:vcorrfunc_cold_hot}) between cold and hot phase velocities for our three standard set of runs, normalised by the rms velocities of the hot and cold phases. At small scales, the hot and cold phase velocities are positively correlated. With increasing separation, the strength of correlation decreases.
	\label{fig:vcf2_cold_hot}
	}
\end{figure}

Spatial correlation functions are another useful tool to understand the coupling between two quantities. We define the correlation function $\mathrm{VCF}_{\mathrm{hc}}(r)$ between the 3D velocity fields of the cold and hot phases as:
\begin{equation}
\mathrm{VCF}_{\mathrm{hc}}(r)=\langle \mathbf{v}_{\mathrm{cold}}(\mathbf{x})\cdot\mathbf{v}_{\mathrm{hot}}(\mathbf{x}+\mathbf{e_1}r)\rangle.    \label{eq:vcorrfunc_cold_hot}
\end{equation}
We choose the pairs $\mathbf{x}$ and $\mathbf{x}+\mathbf{e_1}r$ such that gas at $\mathbf{x}$ is in the cold phase and gas at $\mathbf{x}+\mathbf{e_1}r$ is in the hot phase. 

In \cref{fig:vcf2_cold_hot}, we show the $\mathrm{VCF}_{\mathrm{hc}}(r)$ normalised by the rms velocities of the two phases. For all of our runs, the cold and hot phase velocities are positively correlated at small scales. With increasing separation, the $\mathrm{VCF}_{\mathrm{hc}}(r)$ first increases and reaches a peak. This peak scale may correspond to the cool cloud radius (see the third row of \cref{fig:velocity_projection_maps}), but note that the value of the $\mathrm{VCF}_{\mathrm{hc}}(r)$ at small scales ($r/L<0.04$) may be affected by numerical diffusion.
After this peak, with increasing separation, the correlation becomes weaker. At scales close to the driving scale ($r=L/2$), the velocities are uncorrelated. The coupling between the two phases is weaker for larger $\chi$ for our standard set of runs (f0.10HR, f0.10vwHR and f0.10magHR). It is stronger in the presence of magnetic fields, for our f0.10magHR run (which also has the smallest $\chi$). This trend of stronger coupling between the phases in the presence of magnetic fields is also seen in the cluster scale simulations of \citetalias{Wang2021MNRAS}. For the f0.10hotdriv run, $\mathrm{VCF}_{\mathrm{hc}}(r)$ directly shows the coupling between the phases due to the effect of hydrodynamic drag, which is similar to the $\mathrm{VCF}_{\mathrm{hc}}(r)$ in our fiducial f0.10HR run.

\subsection{Effect of magnetic fields }\label{subsec:magnetic_field_effects}
In the first part of the study, we showed that the $\mathrm{VSF}_2$ in idealised MHD turbulence is closer to the IK scaling, which is shallower than the K41/Goldreich-Sridhar scaling. \cite{Grete2021ApJ} argue that magnetic tension could suppress the kinetic energy cascade and make the power spectra (and $\mathrm{VSF}_2$) shallow, suppressing the bottleneck effect. This is inline with our results.

In our multiphase MHD turbulence runs, we find that magnetic fields couple the hot and cold phases even at the driving scale (see \cref{fig:vsf2_hot_cold_3d}). This efficient coupling is also reflected in stronger spatial correlation between the hot and cold phases in \cref{fig:vcf2_cold_hot}. The hot phase gas has slightly shallower $\mathrm{VSF}_2$ compared to the idealised MHD runs without cooling. But at scales smaller than the driving scale, the $\mathrm{VSF}_2$ of the cold phase gas is much steeper than the hot phase gas. We expect this to be due to the magnetic flux being frozen into the cold phase regions as they condense and contract, as we explained in paragraph \ref{par:vsf2_hc3d_f0.10magHR}. 

\subsection{Effect of projection}\label{subsec:projection_effects}

\begin{figure*}
		\centering
	\includegraphics[width=2\columnwidth]{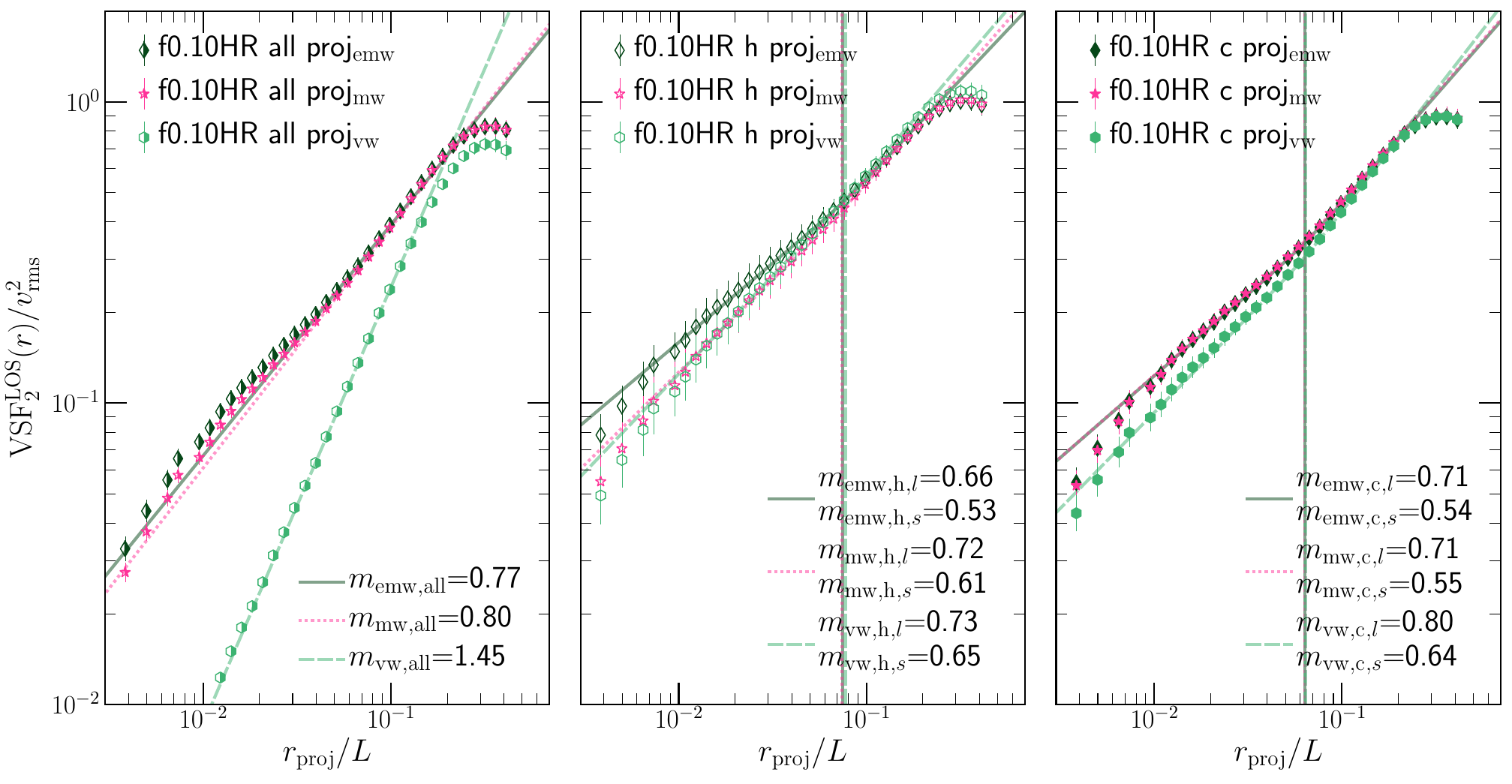}	
	\caption[VSF of with cooling runs: different projection effects]{ The $\mathrm{VSF}_2^{\mathrm{LOS}}$ of the projected velocity field with different weights during projection--emission (emw, $\propto\rho^2$), mass (mw, $\propto\rho$) and volume (vw, $\propto\rho^0$) for our fiducial f0.10HR run. These weights are indicated in the subscripts of the data and fit legends. We show the effects of these weights on the $\mathrm{VSF}_2^{\mathrm{LOS}}$ for gas at all temperatures (first panel), only hot phase gas ($T>10^7$~$\mathrm{K}$, second panel) and only cold phase gas ($T<2\times10^4$~$\mathrm{K}$, third panel). We fit a single power-law to the $\mathrm{VSF}_2^{\mathrm{LOS}}$ in the first panel but broken power-laws to the other two panels. The $\mathrm{VSF}_2^{\mathrm{LOS}}$  shows the maximum cancellation along the LOS and steepening for the volume weighted projection of gas at all temperatures. 
	\label{fig:vsf2_diff_proj_effects}
	}
\end{figure*}

The projection of velocities along the LOS has very different effects on the $\mathrm{VSF}_2$ in the without cooling runs, and the hot and cold phase $\mathrm{VSF}_2$ in the with cooling runs. In part 1, for both hydro and MHD runs, the $\mathrm{VSF}_2^{\mathrm{LOS}}$ becomes steeper by a factor of $r^{0.7}$. 
This is slightly different from the expectations in \cite{Zuhone2016APJ}, who propose a steepening by $r^1$ upon projection. Their result is based on the steepening of the power spectrum by a factor of $k$ upon projection, followed by the conversion between the scaling of the power spectrum to the projected second order structure function. However, we note that the conversion between the scaling of the second order structure function (in real space) and the power spectrum (in Fourier space) is non-trivial. We refer the reader to the appendix of \cite{Stutzki1998A&A} for a derivation of the relation between the scaling of these two quantities for a homogeneous, isotropic field following different power-law spectral indices. However, our turbulence velocity fields do not necessarily follow these assumptions. Moreover, \cite{Esquivel2005ApJ} show that for a Gaussian random field with a steep spectral index $-11/3$ (K41 turbulence has spectral index of $-11/3$), the velocity structure functions steepen less than the factor of $r$ upon projection. This is in agreement with our results. \footnote{\cite{Seta2021inprep} (also see \citealt{Lazarian2008ApJ}) show the importance of using multiple-points or stencils to compute the second order structure function for a field with a steep spectral power law index. We have used a two-point stencil for all our $\mathrm{VSF}_2$ calculations.}

In general, the effect of projection in all our runs depends mainly on the volume-filling fraction of the gas/phase and the uniformity of its emission. The more volume-filling phases have more cancellation upon projection, since larger number of eddies are likely to get cancelled along the LOS during projection. Similarly if the emission is roughly uniform, different eddies along the LOS will have a similar projection weight and are thus more likely to be cancelled upon projection. For the subsonic ($\mathcal{M}\approx0.2$) turbulence without cooling runs, the gas is volume-filling and roughly uniform over the entire volume. So the cancellation effect is strong for the $\mathrm{VSF}_2^{\mathrm{LOS}}$ in the HR and magHR runs. 

Among our runs with cooling, the f0.10vwHR run has uniform volume filling hot phase (see first row of \cref{fig:velocity_projection_maps}), so it shows stronger cancellation of eddies during projection and the $\mathrm{VSF}_2$ steepens by $r^{0.74}$ upon projection. For the fiducial f0.10HR run, the hot phase distribution is scattered and clumpy and the $\mathrm{VSF}_2$ becomes shallow 
on projection. Since most of the cold phase emission occurs from small dense clouds, the cancellation effect of projection is small for the cold phase. 

The distance between two points in the plane of projection $r_{\mathrm{proj}}\leq r$. Upon projection, this effect can shift large scale power to smaller projected scales and make the value of $\mathrm{VSF}_2^{\mathrm{LOS}}$ larger (smaller) at small (large) $r_{\mathrm{proj}}$. It is particularly important for runs with steep 3D $\mathrm{VSF}_2$, such as the cold phase $\mathrm{VSF}_2$ in the f0.10magHR run. Since all 3D $\mathrm{VSF}_2$ are steep at scales smaller than the dissipation scale, this spread of power from large scales to small projected scales can explain the flattening of the $\mathrm{VSF}_2^{\mathrm{LOS}}$ at $r_{\mathrm{proj}}\lesssim0.04$, namely the scales smaller than the dissipation scale.

In \cref{fig:vsf2_diff_proj_effects}, we show the different projected $\mathrm{VSF}_2^{\mathrm{LOS}}$ for the f0.10HR run, where we compare 
different weights on the $v_{\mathrm{LOS}}$ during projection, such as weighted by net emission (emw), weighted by mass (mw) and  weighted by volume (vw). The mw and vw projections are constructed by replacing $\mathcal{L}$ with $\rho$ and 
unity, respectively in \cref{eq:emw-proj-vlos}. We compare 
these three projection weights for different temperature filters - (i) where we consider gas at all temperatures in the first panel, (ii) only consider gas in the hot phase ($T>10^7$~$\mathrm{K}$) in the second panel and (iii) only gas in the cold phase ($T<2\times10^4$~$\mathrm{K}$) in the third panel. We have fit power-laws to the scaling range of the $\mathrm{VSF}_2^{\mathrm{LOS}}$ of all gas and broken power-laws to the scaling range of hot and cold phase $\mathrm{VSF}_2^{\mathrm{LOS}}$. The break scale for the latter two is denoted by a vertical line, similar to \cref{fig:vsf2_hc_projection}. 

In the first panel, for projected $\mathrm{VSF}_2^{\mathrm{LOS}}$ for gas at all temperatures, we observe the strongest variation with the different projection weights. Since the emission and density are larger for the cold dense regions, these projection weights mainly track the small dense clouds and not the volume-filling hot phase. As a result, we do not observe much steepening of the $\mathrm{VSF}_2^{\mathrm{LOS}}$ due to cancellation of eddies. But the volume projection weight mainly tracks the hot and intermediate phase, which 
fill up the volume. This vw projection has more cancellation of eddies along the LOS, especially at smaller scales and shows a much steeper $\mathrm{VSF}_2^{\mathrm{LOS}}$. This steepening due to projection is similar to the turbulence without cooling runs in \cref{fig:nocool_VSF_proj}.

We do not observe very strong effects of changing projection weights for the hot and cold phase $\mathrm{VSF}_2^{\mathrm{LOS}}$. This happens because the volume distribution of both hot and cold phases is clumpy and not volume-filling, as seen in \cref{fig:velocity_projection_maps}. Since the gas phases themselves are not volume-filling, projection along the LOS does not introduce much cancellation. The slopes of the $\mathrm{VSF}_2^{\mathrm{LOS}}$ steepen slightly when we change the projection weight from emw ($\propto \rho^2$) to mw ($\propto \rho$) to vw ($\propto \rho^0$), as the weights become less biased towards the small-sized high density regions and have somewhat more cancellation. Note that we do observe more cancellation for the f0.10vwHR run which has a volume-filling hot phase.

\subsection{Comparison with other studies}\label{subsec:comparison}

\begin{figure}
		\centering
	\includegraphics[width=\columnwidth]{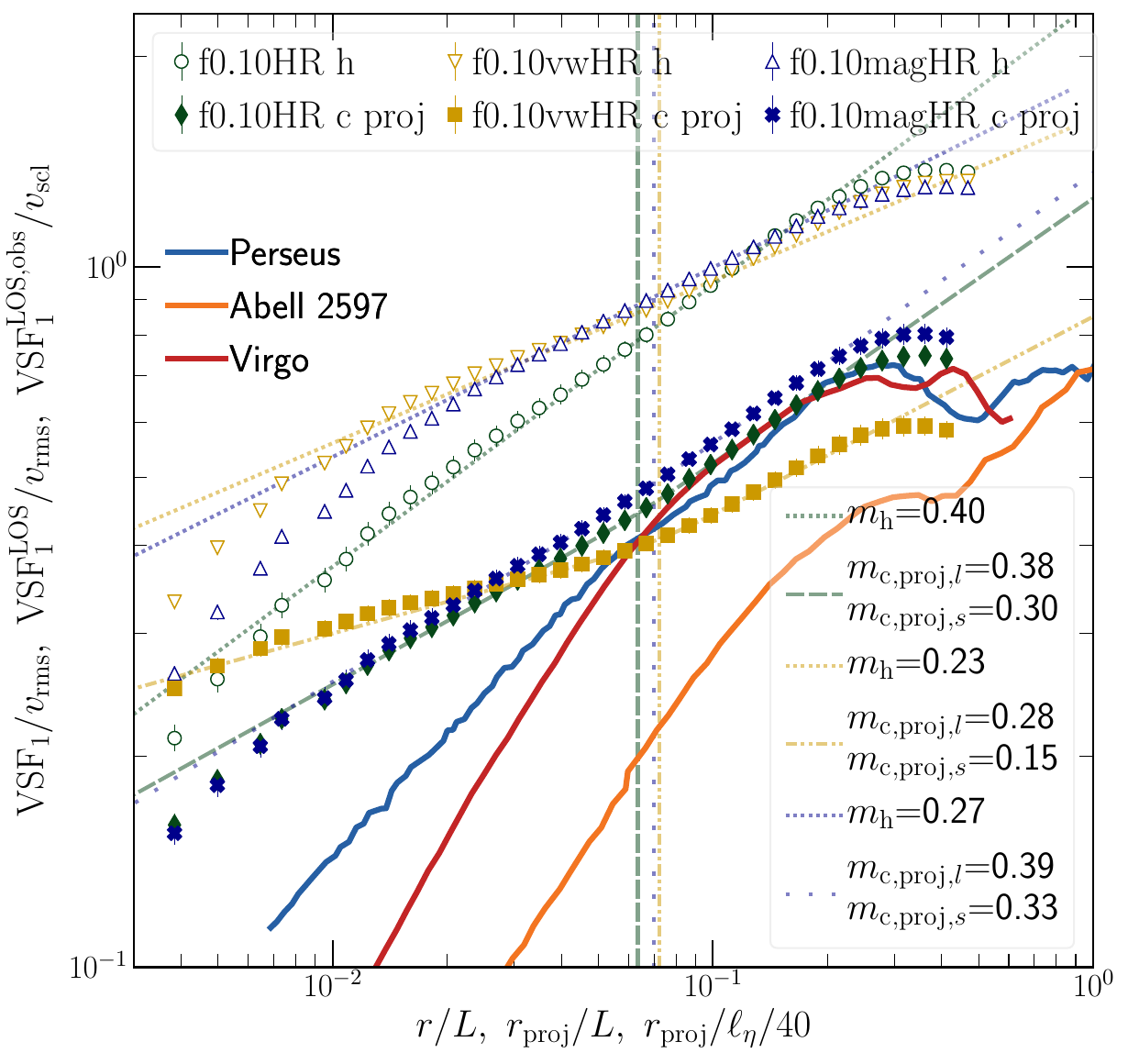}	
	\caption[VSF of with cooling runs: cold projection hot 3d]{ The $\mathrm{VSF}_1$ of the 3D velocity field of the hot phase and the $\mathrm{VSF}_1^{\mathrm{LOS}}$ of the cold phase gas in our different runs. We also show the $\mathrm{VSF}_1^{\mathrm{LOS,obs}}$ from the observations of three clusters from \citetalias{Li2020ApJ}. We have scaled them ($v_{\mathrm{scl}}=200$~$\mathrm{km/s}$) such that they align with $\mathrm{VSF}_1^{\mathrm{LOS}}$ of our cold phase data at the driving scale ($L/2$).  
	The trends across different runs in the hot $\mathrm{VSF}_1$ and cold $\mathrm{VSF}_1^{\mathrm{LOS}}$ for different runs are similar to the trends in hot $\mathrm{VSF}_2$ and cold $\mathrm{VSF}_2^{\mathrm{LOS}}$, respectively. 
	The $\mathrm{VSF}_1^{\mathrm{LOS,obs}}$ are roughly parallel to the cold $\mathrm{VSF}_1^{\mathrm{LOS}}$ at large scales but become steeper at smaller scales.
	\label{fig:vsf1_coldp_hot3d_obs}
	}
\end{figure}
In \cref{fig:vsf1_coldp_hot3d_obs}, we compare 
the cold phase projected first-order velocity structure functions  $\mathrm{VSF}_1^{\mathrm{LOS}}$ ($=(\mathrm{VSF}_{1,x}+\mathrm{VSF}_{1,y}+\mathrm{VSF}_{1,z})/\sqrt{3}$) and the hot phase 3D $\mathrm{VSF}_1$\footnote{Note that we have shown first-order $\mathrm{VSF}$ to compare directly with data from \citetalias{Li2020ApJ}. But we have discussed the second-order $\mathrm{VSF}$s in rest of the paper. }. We also compare the projected cold phase $\mathrm{VSF}_1^{\mathrm{LOS}}$ with the $\mathrm{VSF}_1^{\mathrm{LOS,obs}}$ of H$\alpha$ filaments from \citetalias{Li2020ApJ}. We have imported data from fig.~4 of \citetalias{Li2020ApJ} and have scaled the $\mathrm{VSF}_1^{\mathrm{LOS,obs}}$ by $v_{\mathrm{scl}}=200$~$\mathrm{km/s}$ so that it has similar amplitude as our cold phase $\mathrm{VSF}_1^{\mathrm{LOS}}$. We have also scaled the radial separation ($r_{\mathrm{proj}}/\ell_\eta$) to align the peaks of the Perseus and Virgo data. We have used the same scaling constants for all three clusters. 

The general trends in the hot phase $\mathrm{VSF}_1$ and the cold phase $\mathrm{VSF}_1^{\mathrm{LOS}}$ from our simulations are similar to the trends in their second-order counterparts, which we have discussed in the previous sections. Overall, the cold phase $\mathrm{VSF}_1^{\mathrm{LOS}}$ has a smaller amplitude than the hot phase $\mathrm{VSF}_1$. This can be explained by two main effects - first due to smaller velocity dispersion of the cold phase gas compared to the hot phase (mainly in the two hydro runs) and second due to the shift of power from large scales ($r$) to smaller projected scales ($r_{\mathrm{proj}}$) during projection. At large scales, the slope of the cold phase $\mathrm{VSF}_1^{\mathrm{LOS}}$ is roughly parallel to the hot phase $\mathrm{VSF}_1$ for the f0.10HR run, but it becomes steeper for the f0.10vwHR run and much steeper for the f0.10magHR run. We have discussed the steepening of the cold phase 3D $\mathrm{VSF}_2$ for the f0.10magHR run in  paragraph \ref{par:vsf2_hc3d_f0.10magHR} and \cref{subsec:magnetic_field_effects}. Even though the $\mathrm{VSF}_1^{\mathrm{LOS}}$ flattens strongly due to projection, it is still steeper than the hot phase $\mathrm{VSF}_1$ for this run. All the cold phase projected $\mathrm{VSF}_1^{\mathrm{LOS}}$ flatten at small scales. We have discussed the causes of this flattening in \cref{subsec:projection_effects}. 

Comparing with the observational data from \citetalias{Li2020ApJ}, both $\mathrm{VSF}_1^{\mathrm{LOS,obs}}$ and our $\mathrm{VSF}_1^{\mathrm{LOS}}$ show a peak at large scales. This peak corresponds to the driving scale in our simulations and to the size of the AGN-inflated bubbles in the observational data. At scales larger than the break scale, the $\mathrm{VSF}_1^{\mathrm{LOS}}$ are parallel/slightly shallower than the $\mathrm{VSF}_1^{\mathrm{LOS,obs}}$. 

But at scales smaller than the break scale, $\mathrm{VSF}_1^{\mathrm{LOS}}$ become flatter for all our runs, whereas the $\mathrm{VSF}_1^{\mathrm{LOS,obs}}$ become much steeper and quickly approach zero. 
However, it is important to note that our turbulence with cooling simulations are not convergent with resolution, and the spatial extent of the cold clouds decreases with increasing resolution (see fig.~13 of \citetalias{Mohapatra2021b}, second and third rows). We have discussed the resolution convergence criteria in section 3.6.3 of \citetalias{Mohapatra2021b}. With increasing resolution, the break-scale and the flattening of the projected $\mathrm{VSF}_1^{\mathrm{LOS}}$ are expected to happen at smaller scales due to the smaller cold clouds, which affect the volume distribution of the phases and their projection weight. Increasing the numerical resolution also leads to a smaller viscous scale $\ell_\eta$, which affects the slope of the 3D $\mathrm{VSF}$ (as we showed in \cref{fig:nocool_VSF_convergence}). 

Our projected $\mathrm{VSF}_1^{\mathrm{LOS}}$ do not reproduce the steep slopes in the observed $\mathrm{VSF}_1^{\mathrm{LOS,obs}}$. But in \cref{fig:vsf2_hot_cold_3d}, we showed that the 3D $\mathrm{VSF}_2$ of the cold phase could be strongly magnetised and have a steeper slope than the hot phase $\mathrm{VSF}_2$. The flattening effect of projection could happen at much smaller scales for higher resolution simulations (this is beyond the scope of the present paper). If the slope of the $\mathrm{VSF}_1^{\mathrm{LOS}}$ remains steep upon projection, then magnetic fields can explain the steepening of the cold phase $\mathrm{VSF}_1^{\mathrm{LOS,obs}}$.

\citetalias{Wang2021MNRAS} have studied the $\mathrm{VSF}_1$ of hot and cold phases in a $(1$~$\mathrm{Mpc})^3$ simulation of an isolated galaxy cluster\footnote{The $\mathrm{VSF}_1$ may not be well-converged in a global simulation (see \cref{fig:nocool_VSF_convergence} for our convergence test) and the numerical viscous scale would vary spatially in a refined grid.}. Even though our setup is quite different from that of \citetalias{Wang2021MNRAS}, we observe some similarities between our results. In \citetalias{Wang2021MNRAS}, turbulence in the hot phase is stirred by the motion of the cold phase under the influence of gravity. The $\mathrm{VSF}_1$s of the cold and hot phases are parallel for the hydro runs, and the amplitude of the $\mathrm{VSF}_1$ of the driving phase (in their case the cold phase) is larger.  This is similar to our f0.10hotdriv run, except our hot phase is the driving phase and and the cold phase $\mathrm{VSF}_2$ follows the same scaling as the hot phase $\mathrm{VSF}_2$, but has a smaller amplitude (see \cref{fig:vsf2_hotdriv}). The equivalent scaling of the cold phase $\mathrm{VSF}_1$ with the rest of the ambient gas $\mathrm{VSF}_1$ is also observed in \cite{Gronke2021arXiv} (see their fig.~19), although they start with gas already in a cold cloud and drive turbulence in the ambient hot medium. With the introduction of magnetic fields, in both \citetalias{Wang2021MNRAS} and our study, the velocities of the cold and hot phases become coupled even at large scales.

\cite{Hillel2020ApJ} have analysed the jet-ICM interaction in the central $(40$~$\mathrm{kpc})^3$ region of a cluster. They argue that the steep slope of the $\mathrm{VSF}_1^{\mathrm{LOS,obs}}$ could be due to the interaction of AGN jets and the ICM over a large range of scales. We observe that the  3D $\mathrm{VSF}$s follow a similar scaling relation for both hot and cold phases in our hydrodynamic runs, even when we drive turbulence only in the hot phase (see \cref{fig:vsf2_hotdriv}). Since \cite{Hillel2020ApJ} studied the $\mathrm{VSF}$s only in the hot phase gas, it is possible that the cold phase $\mathrm{VSF}$s would have a similar scaling as the hot phase. However, note that we have mainly studied the effects of steady state turbulence where we inject energy only at driving scales from $10$--$40$~$\mathrm{kpc}$. So we cannot directly compare between the results. A possible way to check this using our setup would be to drive turbulence across a large range of scales and compare the hot and cold phase $\mathrm{VSF}$s, but this is beyond the scope of this study. 

\cite{Kortgen2021MNRAS} have studied the power spectra of column density and projected velocity in a simulation of an isolated galaxy with multiple temperature phases of the interstellar medium. They also identify a break in their projected velocity power spectra and find that it corresponds to a transition from 2D disk to 3D isotropic turbulence for the neutral phases. For the cold phases, the break scale corresponds to the size of the clouds (this is perhaps also true for the peak in our cold-hot VCFs in \cref{fig:vcf2_cold_hot}). 

\section{Caveats and future work}\label{sec:caveats-future}
Here we discuss some of the shortcomings of our work and future prospects. 

One of the main caveats of our setup is the absence of stratification. The ICM is stratified due to the gravitational potential of the dark matter halo. Gravity introduces the free-fall time-scale ($t_{\mathrm{ff}}$), which plays an important role in the ICM thermodynamics \citep{choudhury2016,Choudhury2019,Voit2021ApJ}. For strongly stratified turbulence, the hot phase velocities and their power spectra would be anisotropic, where motion parallel to the direction of gravity is subdued \citep{Mohapatra2020,Mohapatra2021MNRAS} and the eddies form pancake-like structures. On the other hand, the cold clumps, which are 
denser than the hot phase in our hydro runs would have strong motions along the direction of gravity, as seen in \citetalias{Wang2021MNRAS}. The strong X-ray emission from the denser central regions of galaxy clusters may introduce an additional bias in the projected hot phase $\mathrm{VSF}^{\mathrm{LOS}}$. The cancellation effects of projection in real ICM may be small because of stratification. The brightest regions dominate emission along any LOS and cancellation may only be partial. We plan to include stratification and study thermal instability and kinematics of the multiphase gas in a future study.

We have analysed a simple steady state feedback model, with turbulent driving only at large scales and thermal feedback distributing energy uniformly across volume/density. However, the interaction between AGN jets and the ICM could occur across a large range of scales, as shown in \cite{Hillel2020ApJ}. Temporal variability in energy injection due to cool core cycles is another effect that can make the idealised steady state simulations differ from reality (e.g., see fig. 7 of \citealt{prasad2018}). These can also affect the $\mathrm{VSF}^{\mathrm{LOS}}$. Our simple feedback picture is insufficient to capture the physics of other feedback procedures such as cosmic rays \citep{Ji2020MNRAS,Butsky2020ApJ,Butsky2021arXiv}, which are important for both thermal instability models as well as the kinematics of the multiphase halo environments.  

Our simulations are not converged with resolution (this is true for most simulations of the multiphase ICM), as we discussed in section 3.6.3 of \citetalias{Mohapatra2021b}. Higher resolution simulations form more and smaller scale cold gas and have faster cooling rate. Due to our feedback method, this leads to stronger turbulence driving and a cooler hot phase (because of turbulent mixing/heating) upon increasing resolution. Compared to f0.10 run in \citetalias{Mohapatra2021b}, our f0.10HR run has a slightly lower hot phase temperature ($\approx 10^7$~$\mathrm{K}$) and transonic $\mathcal{M}_{\mathrm{hot}}=0.91$. The ICM is observed to be subsonic \citep{hitomi2016}.
With increasing resolution, we also expect the size of the cold clouds to be smaller. These smaller clouds could shift the break scale and the flattening of our projected $\mathrm{VSF}^{\mathrm{LOS}}$ to even smaller scales. 

\section{Conclusions}\label{sec:Conclusion}
In this work we have studied velocity structure functions in homogeneous idealised turbulence and multiphase ICM turbulence. Here are some of our main conclusions:
\begin{itemize}
    \item The 3D $\mathrm{VSF}_2$ of the hot phase gas do not follow the Kolmogorov turbulence scaling ($r^{2/3}$) for any of our turbulence with cooling runs. For our hydro runs, it is steeper than K41 slopes for transonic $\mathcal{M}_{\mathrm{hot}}$ and shallower for subsonic $\mathcal{M}_{\mathrm{hot}}$. For the MHD run, it follows the Iroshnikov-Kraichan scaling ($r^{0.5}$).
    \item The 3D $\mathrm{VSF}_2$ of the cold phase gas has a slightly steeper scaling with $r$ compared to the 3D $\mathrm{VSF}_2$ of the hot phase gas for the hydro runs. It has a smaller amplitude and the ratio of the amplitudes of the cold and hot phase $\mathrm{VSF}_2$ decreases with increasing $\chi$, the ratio of the densities of the cold and hot phases.
    \item For the MHD run, the cold and hot phase gas have similar rms velocity, due to the coupling of the hot and cold phases by the magnetic fields 
    due to magnetic tension. But the cold phase 3D $\mathrm{VSF}_2$ is steeper than the hot phase 3D $\mathrm{VSF}_2$, since the cold phase is magnetically dominated with a low $\beta$ and small scale motions are constrained and subdued.
    \item Upon projection, the projected $\mathrm{VSF}_2^{\mathrm{LOS}}$ follow a broken power-law in the scaling range of the $\mathrm{VSF}_2$. At large scales, the slopes of the projected $\mathrm{VSF}_2^{\mathrm{LOS}}$ are steeper than the 3D $\mathrm{VSF}_2$ slopes only if the corresponding phase is volume-filling to have enough cancellation of velocities during projection. The projected ones are generally shallower than their 3D counterparts at large scales. All the projected $\mathrm{VSF}_2^{\mathrm{LOS}}$ flatten at small scales. 
    \item The projected cold phase $\mathrm{VSF}_1^{\mathrm{LOS}}$ from our simulations show similar trends as the $\mathrm{VSF}_1^{\mathrm{LOS,obs}}$ of H$\alpha$ filaments at large scales. But at small scales, our $\mathrm{VSF}_1^{\mathrm{LOS}}$ flatten due to projection effects, whereas the observational ones do not. This may be a consequence of our limited resolution.
\end{itemize}

\section*{Acknowledgements}
This work was carried out during the ongoing COVID-19 pandemic. The authors would like to acknowledge the health workers all over the world for their role in fighting in the frontline of this crisis. The authors would like to thank the anonymous referee for a constructive report, which helped to improve this work. RM would like to thank Amit Seta for useful discussions. RM acknowledges Prof.~Lisa Kewley and Prof.~Matthew Colless for organising and funding a writing retreat, respectively, where a part of this work was written. PS acknowledges a Swarnajayanti Fellowship (DST/SJF/PSA-03/2016-17) and a National Supercomputing Mission (NSM) grant from the Department of Science and Technology, India. CF acknowledges funding provided by the Australian Research Council (Future Fellowship FT180100495), and the Australia-Germany Joint Research Cooperation Scheme (UA-DAAD). We further acknowledge high-performance computing resources provided by the Leibniz Rechenzentrum and the Gauss Centre for Supercomputing (grants~pr32lo, pr48pi and GCS Large-scale project~10391), the Australian National Computational Infrastructure (grant~ek9) in the framework of the National Computational Merit Allocation Scheme and the ANU Merit Allocation Scheme. The simulation software FLASH was in part developed by the DOE-supported Flash Center for Computational Science at the University of Chicago.

\section{Data Availability}
All the relevant data associated with this article is available upon request to the corresponding author.

\section{Additional Links}
Movies of projected density and temperature of different simulations are available at the following links on youtube:
\begin{enumerate}
    \item \href{https://youtu.be/-_0WIfRL9tI}{Movie} of the f0.10HR simulation.
    \item \href{https://youtu.be/FJnlmVOgQMQ}{Movie} of the f0.10vwHR simulation.
    \item \href{https://youtu.be/NUvp18fKETM}{Movie} of the f0.10magHR simulation.
\end{enumerate}
\section{Software used}
We have used the following software and packages for our work:
FLASH \citep{Fryxell2000,Dubey2008}, matplotlib \citep{Hunter4160265}, cmasher \citep{Ellert2020JOSS}, scipy \citep{Virtanen2020}, NumPy \citep{Harris2020}, h5py \citep{collette_python_hdf5_2014}, LMfit \citep{Newville2016ascl} and astropy \citep{astropy2018}.




\bibliographystyle{mnras}
\bibliography{refs.bib} 



\appendix
\renewcommand\thefigure{\thesection A\arabic{figure}} 
\setcounter{figure}{0}   
\setcounter{table}{0}   
\section*{Appendix A: Density distribution function}\label{app:dens_dist}
\begin{figure}
		\centering
	\includegraphics[width=\columnwidth]{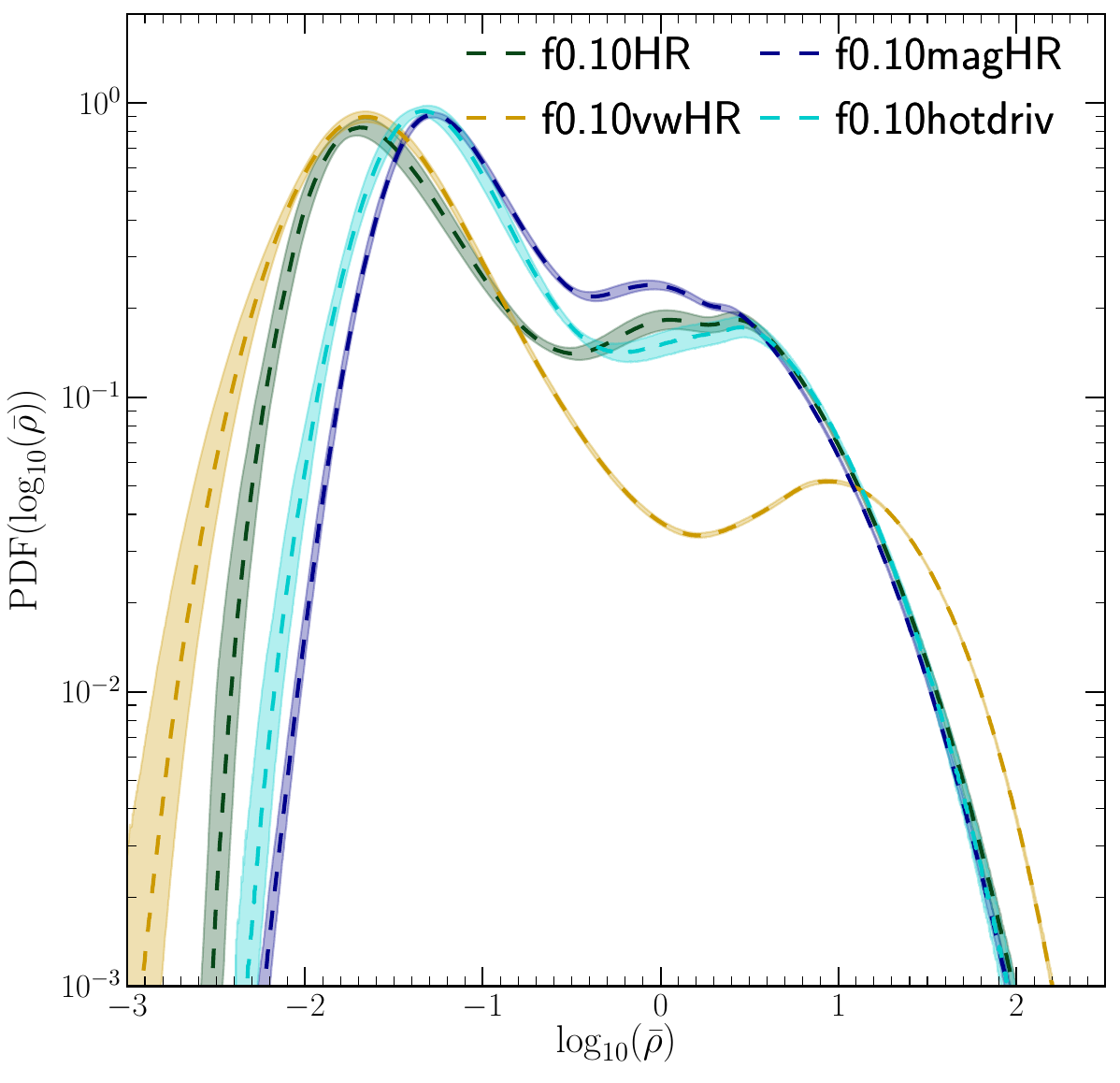}	
	\caption[Density PDF]{The volume PDF of normalised density for our three standard runs and the hot phase driving run. All the PDFs have two peaks, one at small $\bar{\rho}$ corresponding to the hot phase and another at large $\bar{\rho}$ corresponding to the cold phase. The hot and cold phase peaks are the most separated for the f0.10vwHR run, less separated for the f0.10HR run, and closest for the f0.10magHR and f0.10hotdriv runs. 
	\label{fig:dens_volume_PDF}
	}
\end{figure}

In \cref{fig:dens_volume_PDF}, we show the volume PDF of normalised density for our three standard runs and the hot phase driving run. There is a strong contrast in the density of the two phases for the f0.10vwHR run, less for the f0.10HR run, and the least for the f0.10magHR and the hot phase driving runs. 

\section*{Appendix B: structure function slopes and parameters}
\renewcommand\thetable{\thesection B\arabic{table}} 
\begin{table*}
	\centering
	\caption{$\mathrm{VSF}_2$ fit parameters}
	\label{tab:VSF_fit_params}
	\resizebox{\linewidth}{!}{
		\begin{tabular}{cc|c|cccc} 
			\hline
			Label & $v_{\mathrm{rms}}$ & 3D $\mathrm{VSF}_2$ & \multicolumn{4}{c}{Projected $\mathrm{VSF}_2^{\mathrm{LOS}}$}\\
			\hline
			& $(\mathrm{km/s})$ & $m$ & $m_{l}$ & $m_{s}$ & $r_b$  & $\delta$\\
			(1) & (2) & (3) & (4) & (5)& (6) & (7)\\
			\hline
			f0.10HR h & $221\pm15$ & $0.77\pm0.003$ & $0.66 \pm 0.005$ & $0.53 \pm 0.003$ & $0.07 \pm 0.001$ & $0.15\pm0.03$\\
			
			f0.10HR c & $205\pm18$ & $0.86\pm0.001$ & $0.71 \pm 0.002$ & $0.54 \pm 0.003$ & $0.06 \pm 0.001$ & $0.15\pm0.01$\\
			
			f0.10vwHR h & $218\pm12$ & $0.43\pm0.003$ & $1.17 \pm 0.004$ & $0.82 \pm 0.01$ & $0.06 \pm 0.001$ & $0.15\pm0.01$\\
			
			f0.10vwHR c & $167\pm11$ & $0.51\pm0.001$ & $0.53 \pm 0.01$ & $0.26 \pm 0.01$ & $0.07 \pm 0.001$ & $0.15\pm0.03$\\
			
			f0.10magHR h & $185\pm14$ & $0.50\pm0.001$ & $0.70 \pm 0.002$ & $0.65 \pm 0.002$ & $0.07 \pm 0.001$ & $0.13\pm0.02$\\
			
			f0.10magHR c & $183\pm20$ & $1.05\pm0.001$ & $0.73 \pm 0.007$ & $0.62 \pm 0.003$ & $0.08 \pm 0.001$ & $0.15\pm0.02$\\
			\hline
	\end{tabular}}
\end{table*}
In \cref{tab:VSF_fit_params}, we show the fitting parameters for the turbulence with cooling runs. We fit a smoothly broken power-law function `$\mathrm{SBP}$' to the scaling range of the $\mathrm{VSF}_2^{\mathrm{LOS}}$. This function is given by:
\renewcommand\theequation{\thesection B\arabic{equation}} 
\begin{equation}
    \mathrm{SBP}(r,A,r_b,m_s,m_l,\delta)=A\left(\frac{r}{r_b}\right)^{m_s}\left[0.5\left( 1+\frac{r}{r_b}\right)^{1/\delta} \right]^{(m_l-m_s)\delta},\label{eq:sbp_fit}
\end{equation}
where $A$, $r_b$, $m_s$, $m_l$ and $\delta$ are fitting parameters. For $r\ll r_b$, `$\mathrm{SBP}$' scales as $r^{m_s}$ and for $r\gg r_b$, it scales as $r^{m_l}$. The parameter $\delta$ denotes the smoothness of the transition between the two power-laws and has a maximum value of $0.15$. This function is also used in the broken-power law fit in eq.~11 of \cite{Kortgen2021MNRAS}.
\section*{Appendix C: VSF convergence}\label{VSF:convergence}
In \cref{fig:vsf2_convergence}, we show the convergence of the compensated $\mathrm{VSF}_2$ with number of sample pairs of points used to calculate the $\mathrm{VSF}_2$ for a single snapshot at $t=0.911$~$\mathrm{Gyr}$. We use $10^9$ sample pairs for all the 3D $\mathrm{VSF}_2$ computations in the main text. Our algorithm has also been implemented and tested for convergence in \cite{Federrath2008,Federrath2009ApJ,Federrath2021NatAs}.
\renewcommand\thefigure{\thesection C\arabic{figure}} 
\setcounter{figure}{0}  
\begin{figure}
		\centering
	\includegraphics[width=\columnwidth]{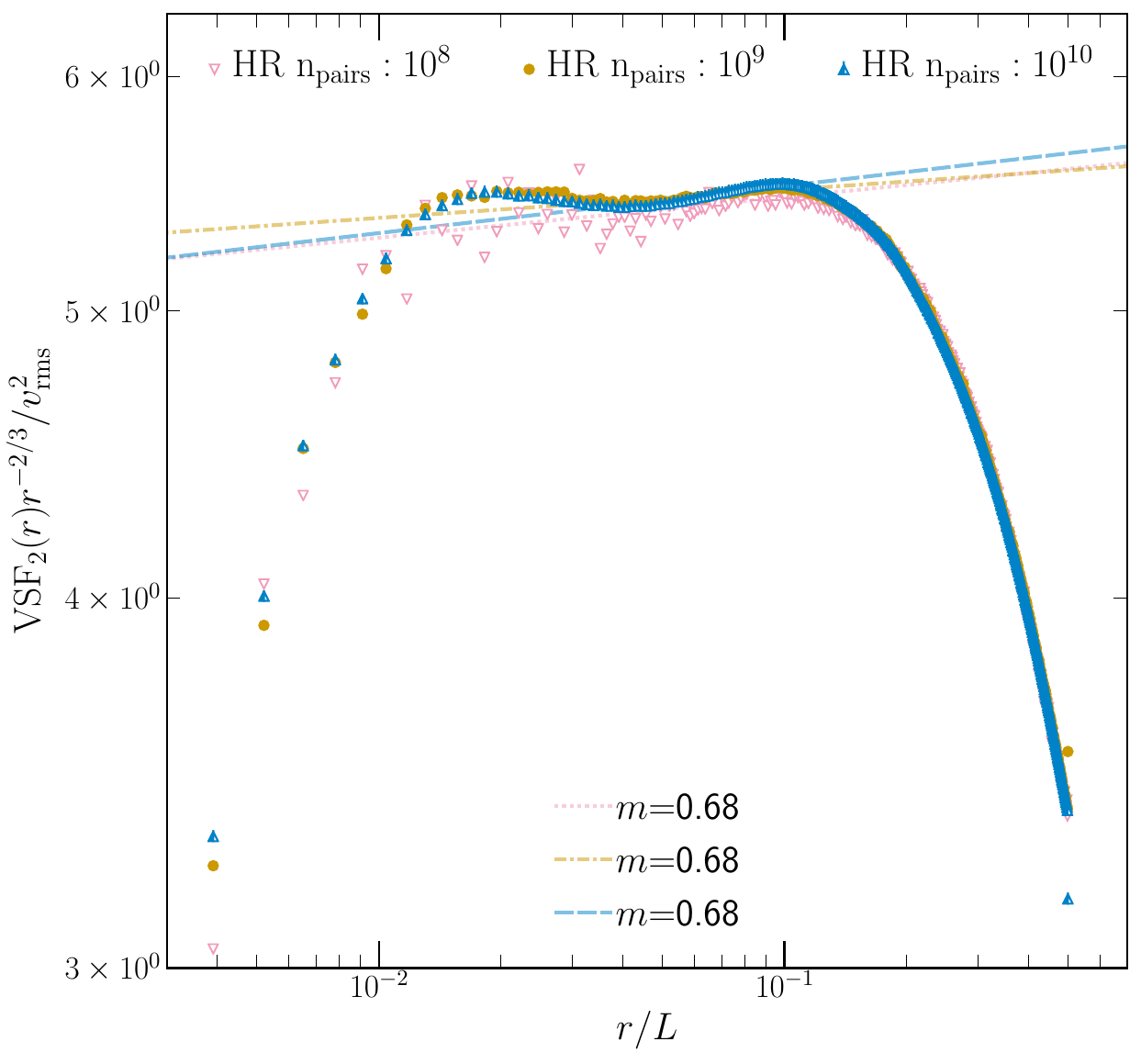}	
	\caption[VSF convergence]{The compensated $\mathrm{VSF}_2$ of the HR runs ($768^3$ resolution without cooling run) for three different sample number of pairs of points ($n_{\mathrm{pairs}}$).  We obtain convergence for $n_{\mathrm{pairs}} = 10^9$, and choose this to evaluate the $\mathrm{VSF}_2$ in this study. 
	\label{fig:vsf2_convergence}
	}
\end{figure}

\bsp	
\label{lastpage}
\end{document}